\makeatletter
\@ifundefined{@parse@version@dash}{%
\def\@parse@version#1{\@parse@version@0#1}
\def\@parse@version@#1/#2/#3#4#5\@nil{%
\@parse@version@dash#1-#2-#3#4\@nil}
\def\@parse@version@dash#1-#2-#3#4#5\@nil{%
  \if\relax#2\relax\else#1\fi#2#3#4 }
}{}
\makeatother

\documentclass[aps,prb,reprint,amsmath,amssymb]{revtex4-2}
\usepackage{graphicx}
\usepackage{xcolor}

\renewcommand\vec[1]{{\bf #1}}
\newcommand{\ket}[1]{|#1 \rangle}
\newcommand{\bra}[1]{\langle #1|}
\newcommand\modif[1]{}

\begin{document}

\title{Modeling of spin decoherence in a Si hole qubit perturbed by a single charge fluctuator}

\author{Baker Shalak}
\affiliation{Universit\'e de Lille, CNRS, Centrale Lille, Universit\'e Polytechnique Hauts-de-France, Junia, UMR 8520 - IEMN, F-59000 Lille, France}
\author{Yann-Michel Niquet}
\affiliation{Univ. Grenoble Alpes, CEA, IRIG-MEM-L$\_$Sim, F-38000, Grenoble, France}
\author{Christophe Delerue}
\email{christophe.delerue@iemn.fr}
\affiliation{Universit\'e de Lille, CNRS, Centrale Lille, Universit\'e Polytechnique Hauts-de-France, Junia, UMR 8520 - IEMN, F-59000 Lille, France}

\begin{abstract}
Spin qubits in semiconductor quantum dots are one of the promising devices to realize a quantum processor. A better knowledge of the noise sources affecting the coherence of such a qubit is therefore of prime importance. In this paper, we study the effect of telegraphic noise induced by the fluctuation of a single electric charge. We simulate as realistically as possible a hole spin qubit in a quantum dot defined electrostatically by a set of gates along a silicon nanowire channel.
Calculations combining Poisson and time-dependent Schr\"odinger equations allow us to simulate the relaxation and the dephasing of the hole spin as a function of time for a classical random telegraph signal. We show that dephasing time $T_2$ is well given by a two-level model in a wide range of frequencies. 
Remarkably, in the most realistic configuration of a low frequency fluctuator, the system has a non-Gaussian behavior in which the phase coherence is lost as soon as the fluctuator has changed state. The Gaussian description becomes valid only beyond a threshold frequency $\omega_{th}$, when the two-level system reacts to the statistical distribution of the fluctuator states. We show that the dephasing time $T_{2}(\omega_{th})$ at this threshold frequency can be considerably increased by playing on the orientation of the magnetic field and the gate potentials, by running the qubit along "sweet" lines.
However, $T_{2}(\omega_{th})$ remains bounded due to dephasing induced by the non-diagonal terms of the stochastic perturbation Hamiltonian.
On the other hand, our simulations reveal that the spin relaxation, usually characterized by the time $T_1$, cannot be described cleanly in the two-level model because the coupling to higher energy hole levels impacts very strongly the spin decoherence. 
This result suggests that multi-level simulations including the coupling to phonons should be necessary to describe the relaxation phenomenon in this type of qubit.

\end{abstract}

\maketitle

\section{Introduction}

Spin qubits are being actively studied for quantum computing \cite{Kane98,Loss98}. One path that is being particularly explored at the moment is the use of silicon or germanium qubits \cite{Pla12,Zwanenburg13,Kawakami14,Veldhorst15,Takeda16,Yoneda18,Zajac18,Huang19}, as it promises extreme miniaturization and integration while benefiting from the expertise and resources of microelectronic technologies. The use of isotopically purified Si substrates also allows one, by suppressing the hyperfine interaction between electrons and nuclear spins, to obtain very long electron spin lifetimes on donors \cite{Tyryshkin12} and in quantum boxes defined by electrostatic confinement \cite{Veldhorst14}. This lifetime is particularly long for electrons in conduction band states due to the weak spin-orbit coupling \cite{Zwanenburg13} but this makes the manipulation of electron spin via electrical signals not very efficient \cite{Corna18,Bourdet18}.

In this context, hole qubits receive growing interest because of the stronger spin-orbit coupling in the valence band allowing efficient manipulation of the effective spin by electrical means \cite{Bulaev05,Bulaev07,Kloeffel13,Li15,Maurand16,Hendrickx20}. Recent work has demonstrated Rabi oscillations with frequencies of several hundred MHz in silicon and germanium hole qubits \cite{Maurand16,Crippa18,Crippa19,Watzinger18,Hendrickx20}. Two qubit gates have been realized recently \cite{Veldhorst15,Watson18,Zajac18,Xue19,Huang19,Hendrickx20_2}. In addition, the strong interaction between spin and microwave photons makes long distance coupling between qubits possible \cite{Abadillo21,Holman21,Ibberson21,Michal22,Yu22,Bosco22}. However, the effective spin-orbit coupling depends on the spatial profile of the hole wave function, lattice deformations and electric fields, which increases the variability between devices \cite{Martinez22} and makes the qubits much more sensitive to phonons and electric potential fluctuations \cite{Culcer09,Bermeister14,Yoneda18}. It is therefore essential to better understand the influence of these phenomena on the coherence lifetimes of spins in hole qubits in silicon technology. Recent theoretical works have focused on the spin-phonon coupling \cite{Maier13,Li20}, we are interested here in the influence of charge fluctuations which is usually dominant at low temperature.

Many theoretical studies have investigated the nature and strength of spin-orbit coupling in the heavy hole, light hole and split-off states of the valence band \cite{Kloeffel13,Kloeffel18}. Proposals have also been made to minimize the effects of electric potential fluctuations \cite{Paladino14}, to find operating points (the so-called "sweet" spots) where the Larmor frequency becomes insensitive to the fluctuations  \cite{Venitucci18,Venitucci19,Benito19,Bosco21}. The considerable increase in the hole spin coherence time at such sweet spots has actually been demonstrated recently in a silicon-on-insulator (SOI) device \cite{Piot22}.

Our goal in this paper is different, it is to better understand the physics of hole spin decoherence under the effect of charge fluctuations in a device that is as realistic as possible compared with what was realized experimentally. We consider the case of a hole qubit made on an SOI and formed by electrostatic confinement within a silicon nanowire \cite{Maurand16,Crippa18}. This qubit is subjected to telegraphic noise due to the fluctuation of a single charge between a metal gate and its neighboring oxide. 

The telegraphic noise can be seen as a minimal model reproducing main features of $1/f$ noise \cite{Bergli09,Paladino14}. Here we are not interested in the action of a large number of fluctuators leading to a $1/f$ noise but we aim to better understand the effect of a single one on the qubit. The evolution of the electronic states of the qubit as a function of time is calculated by numerical solution of the Schrödinger equation in a multi-band framework in which the potential is calculated taking into account the complex environment of the qubit. Such a description is necessary because of the strong sensitivity of the spin-orbit coupling to the potential profile in the vicinity of the hole wave function. 

In this paper, we compare the results of numerical calculations with analytical models from the literature that have been established for a two-level system coupled to telegraphic noise, allowing us to understand the evolution of the relaxation time $T_1$ and dephasing time $T_2$ as a function of noise intensity and tunneling rate $\nu$. However, we show that the potential fluctuations couple with more than two levels, and that it is necessary to take a large number of them ($\sim 20$) to simulate the spin evolution as a function of time. This leads us to define another decoherence time $T_{1}'$ that represents the mean time the system remains describable as a two-level system. Despite this complexity, we obtain that, for experimentally relevant tunneling rates $\nu$ below a certain value $\omega_{th}$, $T_2$ has a remarkable but very simple behavior, being equal to $2/\nu$ independently of the intensity of the perturbing potential. 
We also study how the operating conditions (gate bias, magnetic field orientation) affect the dephasing time $T_{2}(\omega_{th})$ at the threshold frequency $\omega_{th}$.
In addition, we obtain that the decoherence time $T_{1}'$ is always smaller than the relaxation time $T_1$ calculated in the two-level model.
This reveals that the spin relaxation dynamics is very strongly influenced by the coupling with hole states further away in energy, and thus $T_1$ cannot be described simply on the basis of the two-level model, a conclusion that should guide future theoretical simulation work.

\section{Methodology}

\subsection{Calculation of the potential and the hole states in the device}

\begin{figure}%[!t]
\centering
\includegraphics[width=0.9\columnwidth]{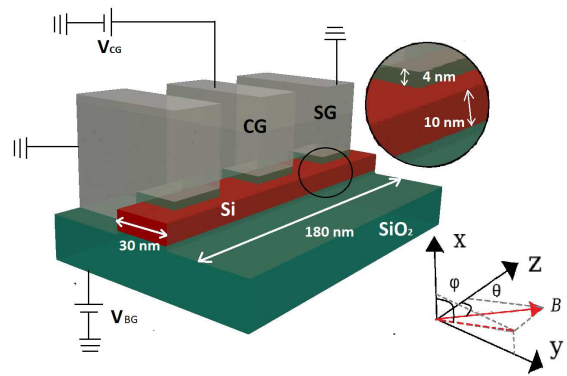}
\caption{Schematics of the hole qubit device consisting of a 10 nm thick Si nanowire channel (red) on top of a buried oxide (green). Top gates (gray) partly cover the nanowire (over 20~nm for a total width of the nanowire of 30~nm). The gate stack is made of SiO$_2$ (green). CG represents the central gate that defines the hole quantum dot. SG is the secondary gate discussed in this paper. The orientation of the magnetic field $\mathbf{B}$ is characterized by the polar angle $\theta$ and azimuthal angle $\varphi$ defined in the figure.}
\label{3Ddevice}
\end{figure}

The device presented in Fig.~\ref{3Ddevice} is a Metal-Oxide-Semiconductor Field Effect Transistor (MOSFET) formed by a Si nanowire oriented along [110] (hereafter, the $z$ axis). The nanowire has a rectangular section with width (along $y$) of 30~nm [lateral $(1\overline{1}0)$ facets] and thickness of 10~nm [$(001)$ facets] and is lying on a 25-nm thick SiO$_2$ buried oxide deposited on a doped Si substrate which can be used as a back gate.  On top of the channel, there are metal gates with length and separation along $z$ of 30~nm that partly envelop the channel (over 20~nm). A 4-nm thin layer of SiO$_2$ separates the metallic gates from the nanowire. The transistor is covered with Si$_3$N$_4$.
The central gate (CG) is used to fix the potential that will induce the formation of a quantum dot (with corner states) in the nanowire \cite{Venitucci18}.
Two secondary gates are arranged along the $z$ axis, to the right and left of the central gate. The central gate is biased at $V_{CG}=-0.1$~V and the other gates are grounded in order to confine the hole in the central quantum dot. 
A static magnetic field $\mathbf{B}$ is applied along a direction defined by polar $(\theta)$ and azimuthal $(\varphi)$ angles (see Fig.~\ref{3Ddevice}).

The potential induced by the gates or by the presence of a charge impurity in the oxide layer is calculated by solving the Poisson equation linking the charge density $\rho$ and the dielectric constant $\epsilon$ that both depend on the position. To solve it, we use the finite difference method which consists in discretizing the equation spatially on a 3D mesh. 

\begin{figure}%[!h]
\centering
\includegraphics[width=0.9\columnwidth]{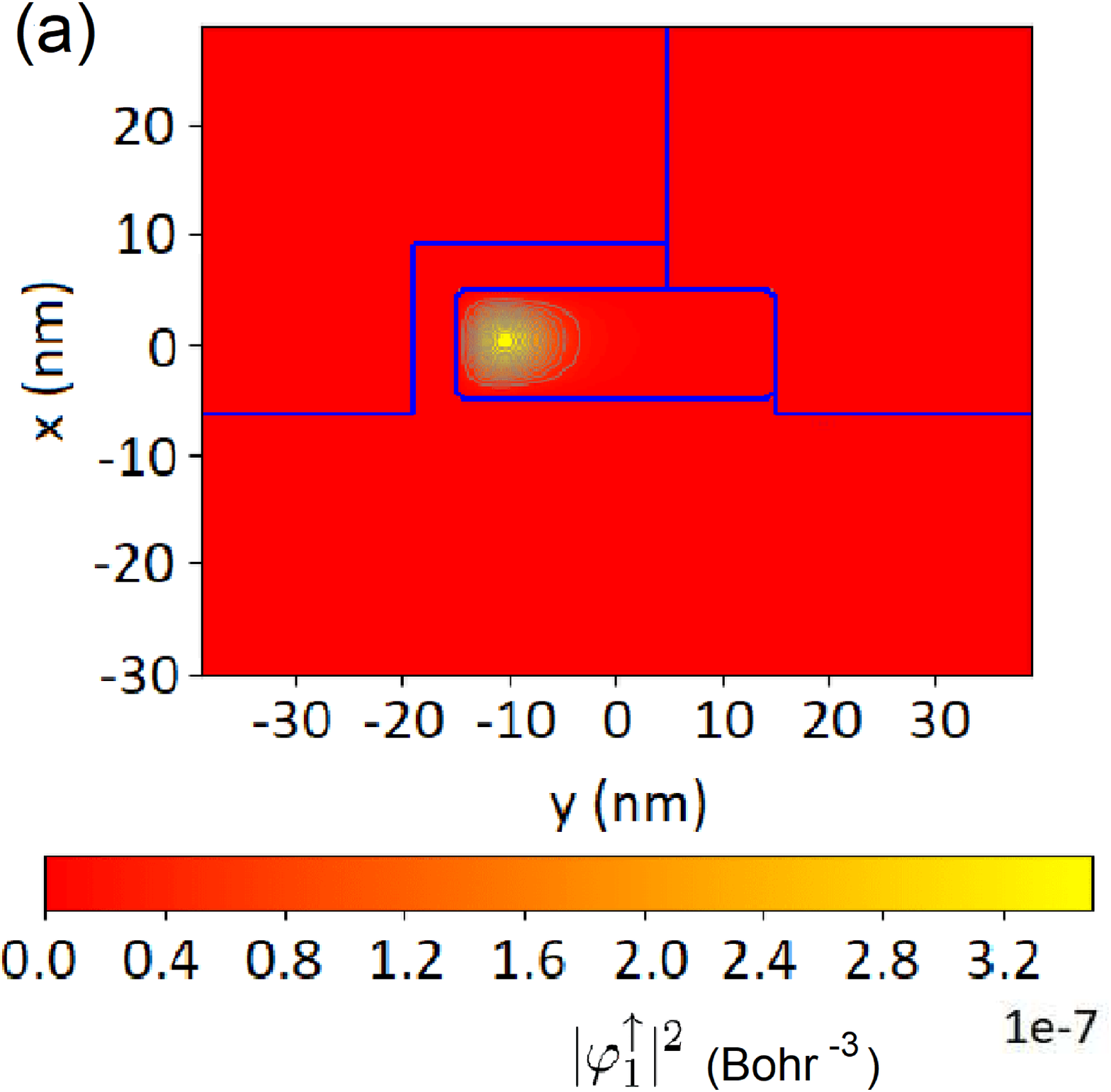}
\includegraphics[width=0.9\columnwidth]{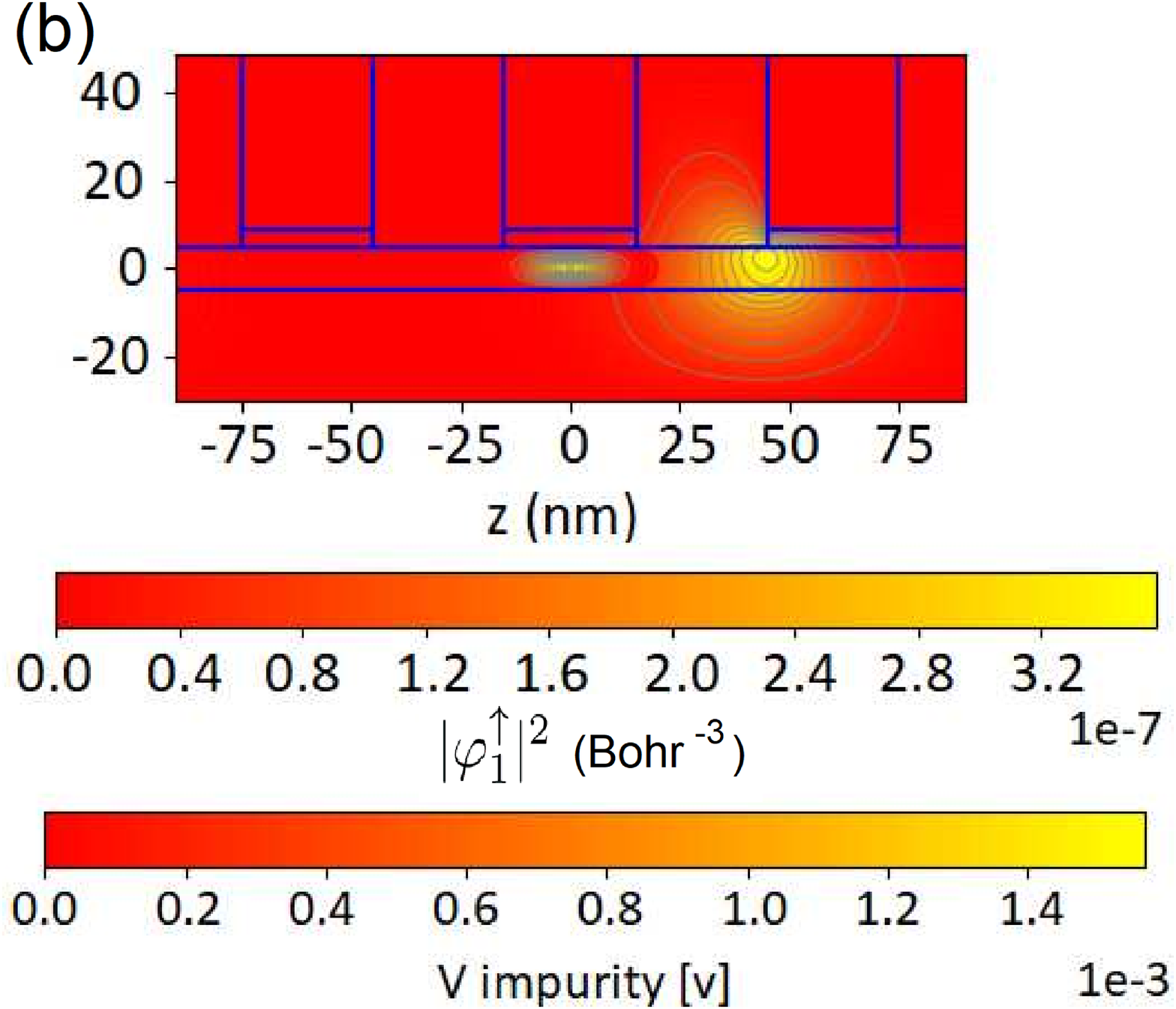}
\caption{Iso-density surface of the ground-state hole wave function depicted across (a) transverse [$xy$ plane at $z=0$] and (b) longitudinal [$xz$ plane, $y=-10$~nm corresponding to the center of the hole wave function] cross sections. The potential induced by a single charge (Trap 3) at position $x=8.4$~nm, $y=0$ and $z=46.0$~nm is also shown in panel (b).}
\label{fig_wf_pot}
\end{figure}

%\subsection{using K.P method to calculate the W.F states}
To calculate the electronic structure in our device, we use a six bands $\mathbf{k} \cdot \mathbf{p}$ model which gives an excellent description of the valence band states, including the effect of spin-orbit coupling. Details on the numerical methods are given in Ref.~\cite{Venitucci18} and are reproduced in Sect.~I of the Supplemental Material \cite{supplemental} for convenience. 
The wave functions of the holes are written as

\begin{equation}
\psi(\vec{r}) = \sum_{\alpha} F_{\alpha}(\vec{r}) u_{\alpha}(\vec{r})
\end{equation}

\noindent where $F_{\alpha}(\vec{r})$ is an envelope function and $u_{\alpha}(\vec{r})$ is a Bloch function in the set $\{\ket{\frac{3}{2},+\frac{3}{2}},\ket{\frac{3}{2},+\frac{1}{2}},\ket{\frac{3}{2},-\frac{1}{2}},\ket{\frac{3}{2},-\frac{3}{2}},\ket{\frac{1}{2},+\frac{1}{2}},\ket{\frac{1}{2},-\frac{1}{2}}\}$ .
The envelope functions are solutions of six coupled differential equations obtained from the $\mathbf{k} \cdot \mathbf{p}$ Hamiltonian $H_{6kp}$ given in Sect.~I of the Supplemental Material \cite{supplemental} in which the wavevector $\vec{k}$ has been substituted by $-i\vec{\nabla}$. These equations are discretized on a finite difference mesh. Even if the quantum dots are effectively decoupled by the action of the lateral gates, periodic boundary conditions are applied along $z$. The surface of the wire is considered as a hard wall for the wavefunction. The effect of the potential vector $\vec{A}$ on the envelope functions is included through Peierls's substitution \cite{Vogl95}. The effect of the magnetic field on the Bloch functions is described by the following Hamiltonian \cite{Luttinger56}:

\begin{equation}
H_{\rm Bloch}=-(3\kappa+1)\mu_B\vec{B}\cdot\vec{L}+g_0\mu_B\vec{B}\cdot\vec{S}=\mu_B\vec{B}\cdot\vec{K}\,,
\label{eqHbloch}
\end{equation}

\noindent where $\vec{L}$ is the (orbital) angular momentum of the Bloch function, $\vec{S}$ is its spin, and $\kappa=-0.42$ in silicon. The expression of the matrices \vec{K} is given in Sect.~I of the Supplemental Material \cite{supplemental}.

The hole qubit states are taken as the topmost valence band states. 
The wave function of the highest hole state is presented in Fig.~\ref{fig_wf_pot}a across a transverse section of the MOSFET and in Fig.~\ref{fig_wf_pot}b for a longitudinal one.

\subsection{Fluctuator model and time dependent Hamiltonian}

\begin{figure}%[!h]
\centering
\includegraphics[width=0.8\columnwidth]{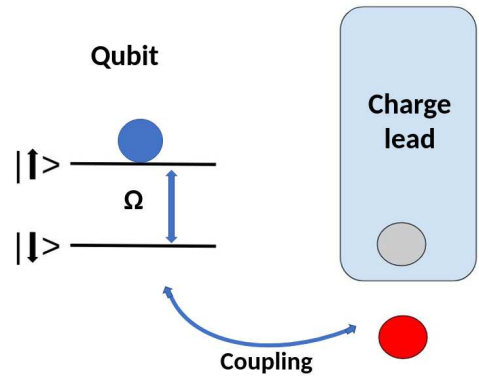}
\caption{Model of qubit coupled to a single charge fluctuator. A single electron tunneling between a charge lead, i.e., a gate, and a point trap (red circle) induces a sudden change in the electrostatic potential that couples to the qubit hole states. Dephasing and relaxation of the hole spin take place under the effect of this perturbation in the form of a telegraphic noise. Only the two hole levels of lowest energy are depicted for simplicity. The Zeeman energy splitting between them is $\hbar\Omega$ in which $\Omega$ is the Larmor angular frequency.}
\label{model}
\end{figure}

We consider that the hole qubit is coupled to a charge fluctuator (Fig.~\ref{model}) which follows a random telegraphic signal $\chi(t)$ that describes the filling of a localized charge trap in the oxide layer at a distance of 1~nm from a metallic gate, either the central gate (CG) or the secondary gate (SG) shown in Fig.~\ref{3Ddevice}. 
An example of a potential created by a localized charge $-e$ under the secondary gate is presented in Fig.~\ref{fig_wf_pot}(b).
The charge fluctuator is described as a random telegraphic noise \cite{Paladino14}, i.e., $\chi(t)$ takes two values, 0 or 1, with respective probabilities $p_{0}$ and $p_{1}$. In the state $0$ of $\chi(t)$, the trap is empty, and in the state $+1$, a charge $-e$ has tunnelled from the gate to the trap with a transition rate $\nu_{0 \to 1} = \nu [1-f_{FD}(\varepsilon_{0})]$ where $\nu$ is the tunneling rate, $f_{FD}$ is the Fermi-Dirac distribution function and $\varepsilon_{0}$ is the position of the trap level with respect to the Fermi level in the reservoir (gate). Here, we assume for simplicity $\varepsilon_{0}=0$, $\nu_{0 \to 1} = \nu/2$, $\nu_{1 \to 0} = \nu - \nu_{0 \to 1} = \nu/2$, $p_{0} = p_{1}= 1/2$ \cite{abelthese}. 
In this model, the "classical" frequency of the  telegraphic signal, i.e. the average number of switches per time unit, is given by $\nu_{cl}=\nu/2$ .

The time-dependent Hamiltonian of the system reads as

\begin{equation}
H(t)= H_{0}+ \chi(t) U
\end{equation}

\noindent where $H_{0}$ is the hamiltonian representing the system under the static magnetic field $\mathbf{B}$ but without any electrical perturbation.
$U$ defines the perturbation when a charge is on the trap.
  
In order to calculate the evolution of the wave function $|\psi(t)\rangle = \exp (-i/\hbar \int_{0}^{t}H(t')dt') |\psi(0)\rangle$ and the characteristic times $T_{1}$, $T_{1}'$ and $T_{2}$ which will be discussed in the next sections, we solve the time-dependent Schr\"{o}dinger equation numerically.
A Chebyshev polynomial expansion is used to describe the time propagation of the wave function \cite{Leforestier91}. This approach ensures high numerical stability of the propagator.

\subsubsection{Two-level model}

A qubit is generally constructed as a two-level system, the two states corresponding here to the lowest energy hole states, eigenstates of $H_0$, denoted as $|\varphi_{1}^{\uparrow} \rangle$ and $|\varphi_{1}^{\downarrow} \rangle$ in reference to spin-$1/2$ systems.
The Zeeman splitting $(\propto B)$ between the two levels (for $\chi(t)=0$) is written as $\hbar\Omega$ where $\Omega$ is the Larmor angular frequency.
Rabi oscillations between the two states can be electrically driven by a radio-frequency signal with an angular frequency close to $\Omega$ on the central front gate \cite{Venitucci18,Li20}.

The matrices of the hamiltonian and the perturbation in this basis set are given by

\begin{equation} \label{huzero}
H_{0}= \frac{\hbar}{2} \left(
\begin{array}{cc}
\Omega &  0 \\ 
0 & -\Omega 
\end{array} \right)
\quad 
U= \left(
\begin{array}{cc}
u_{\uparrow\uparrow} &  u_{\uparrow\downarrow} \\ 
u_{\uparrow\downarrow}^{*} & u_{\downarrow\downarrow} 
\end{array} \right)
\end{equation}

The electrostatic potential induced by the trapped charge does not explicitly involve spin but the matrix elements of $U$ depend on $B$ through $|\varphi_{1}^{\uparrow} \rangle$ and $|\varphi_{1}^{\downarrow} \rangle$. 
The coupling terms between these opposite spin states result from spin-orbit coupling and time-reversal symmetry breaking under the effect of the magnetic field $\mathbf{B}$.
As shown in Sect.~II of the Supplemental Material \cite{supplemental}, the matrix elements of $U$ behave as 

\begin{eqnarray}
&&u_{\uparrow\downarrow} = \eta_{\uparrow \downarrow}(\vec{b}) B \nonumber \\
&&u_{\uparrow\uparrow} = u_{0} + \eta_{\uparrow \uparrow}(\vec{b}) B \nonumber \\
&&u_{\downarrow\downarrow} = u_{0} + \eta_{\downarrow \downarrow}(\vec{b}) B \label{eq_U_elements}
\end{eqnarray}

\noindent in which $\vec{b}=\vec{B}/B$ and therefore $\eta_{\uparrow \downarrow}$, $\eta_{\uparrow \uparrow}$ and $\eta_{\downarrow \downarrow}$ just depend on the orientation of $\vec{B}$. $u_{0}$ is a rigid shift of the two energy levels under the effect of the perturbation. 

The interaction of the qubit with its environment (Fig.~\ref{model}) causes a loss of information, called decoherence. It is usually separated in two processes, relaxation of characteristic time $T_{1}$ and dephasing of characteristic time $T_{2}$ \cite{Galperin04,Paladino14}.
We obtain $T_{1}$ and $T_{2}$ by calculating the evolution with time of $\langle\langle \sigma_{i}(t) \rangle\rangle = \langle \psi(t) | \sigma_{i} | \psi(t) \rangle_{\{E\}}$ with $i=1,2,3$. $\sigma_{1}$, $\sigma_{2}$ and $\sigma_{3}$ are the $2 \times 2$ Pauli matrices written in the basis of $|\varphi_{1}^{\uparrow} \rangle$ and $|\varphi_{1}^{\downarrow} \rangle$ (they are not written $\sigma_{x}$, $\sigma_{y}$ and $\sigma_{z}$ since $x$, $y$ and $z$ refer to the geometrical axes of system).
The subscript $\{E\}$ means that an average is taken over many (1000) realizations of the telegraph noise, i.e. of the environment.

The relaxation is the loss of information by the process $|\varphi_{1}^{\uparrow} \rangle \leftrightarrow |\varphi_{1}^{\downarrow} \rangle $ due to the stochastic variations of the non diagonal term $\chi (t) u_{\uparrow\downarrow}$ of the hamiltonian.  Starting with the condition $|\psi(0)\rangle = |\varphi_{1}^{\uparrow}\rangle$, we calculate $T_1$ by fitting with an exponential function the decay of $\sigma_{\parallel}(t) = \langle\langle \sigma_{3}(t) \rangle\rangle$ over time.

The dephasing comes from the changes $\delta \phi(t)$ of the phase characterizing the spin precession due to the stochastic variations of the terms of the hamiltonian, $\chi (t) U$.
Indeed, in the $\chi(t)=1$ state, the Larmor angular frequency changes to $\Omega'$ where $\hbar\Omega'$ is the Zeeman splitting obtained by diagonalization of $H_{0}+U$.
As discussed in Sect.~III of the Supplemental Material \cite{supplemental}, it will be interesting to define the (threshold) angular frequency

\begin{equation}
\omega_{th}= |\Omega-\Omega'|
\end{equation}

\noindent that characterizes the change of phase velocity (usually $\omega_{th} \ll \Omega$). 
$\hbar \omega_{th}$ represents the change in the Zeeman splitting between the two states of the fluctuator. 
We deduce from Eq.~(\ref{eq_U_elements}) that $\omega_{th} \propto B$ in most cases (see Sect.~III.B of the Supplemental Material \cite{supplemental}). 

A measure of the phase coherence is given by the quantity $\left\langle \exp (i \delta \phi(t) ) \right\rangle_{\{E\}}$ \cite{Paladino14}.
Equivalently, we have calculated the quantity $m(t) = \left| \langle\langle \sigma_{1}(t) \rangle\rangle + i \langle\langle \sigma_{2}(t) \rangle\rangle \right|$ using the initial condition $|\psi(0)\rangle = (|\varphi_{1}^{\uparrow}\rangle + |\varphi_{1}^{\downarrow}\rangle)/\sqrt{2}$.
The decay of $m(t)$ from 1 to 0 comes from the dephasing between the different realizations of the potential fluctuations.   
$T_{2}$ is obtained by fitting with an exponential function the decay of $m(t)$ over time.
It is important to note that for $\nu < \omega_{th}$, $m(t)$ exhibits damped oscillations at a frequency of the order of $\omega_{th}$ \cite{Paladino14} (see Sect.~VIII of the Supplemental Material \cite{supplemental}). In this case, $T_2$ is obtained from the exponential decay of the envelope.

\subsubsection{Multi-level model}

The perturbation generated by the fluctuator induces coupling terms that are not limited to the two states considered above. We have therefore considered a model integrating $2N$ hole states. With the matrices of $H_0$ and $U$ written in this basis, we compute the propagation of the hole wave function as a function of time starting from the same initial conditions. We deduce the observable $m(t)$ from which we obtain the characteristic time $T_2$, assuming that $\boldsymbol{\sigma}$ acts only in the subspace formed by $|\varphi_{1}^{\uparrow} \rangle$ and $|\varphi_{1}^{\downarrow} \rangle$. As a matter of fact, during the evolution as a function of time, the weight of the wave function of the hole on the two initial states, $p_{1}(t) = \left|\langle \varphi_{1}^{\uparrow} | \psi(t) \rangle \right|^{2} + \left|\langle \varphi_{1}^{\downarrow} | \psi(t) \rangle \right|^{2}$, decreases under the effect of the couplings to the other states. From the exponential decay of $\left\langle p_{1}(t) \right\rangle_{\{E\}}$, averaged over all the realizations of the telegraphic noise, we deduce another decoherence time, which we call $T_{1}'$, following the methodology described in Appendix~\ref{appendix_cal_t1}.

It is important to note that the elements composing the perturbation $U$ are not independent since they are matrix elements of the same electrostatic potential. The same effect is at the origin of all the decoherence mechanisms considered here. 

\subsubsection{Time interval, frequency range and trap position}

We consider (except where otherwise stated) a magnetic field of 0.2712~T oriented along the direction characterized by $\theta=90^{\circ}$ and $\varphi=45^{\circ}$ (Fig.~\ref{3Ddevice}) which leads to a Larmor frequency $\Omega/(2\pi)$ of 10~GHz. This forces us to use a time step of $10^{-12}$~s for the numerical solution of the time dependent Schr\"{o}dinger equation for $\nu \leq 2 \times 10^{11}$~s$^{-1}$, $10^{-13}$~s for $\nu = 2 \times 10^{12}$~s$^{-1}$, $10^{-14}$~s for $\nu = 2 \times 10^{13}$~s$^{-1}$ and $10^{-15}$~s for $\nu = 2 \times 10^{14}$~s$^{-1}$. The maximum simulation time has been limited to $10^{-4}$~s. We thus considered $\nu$ between $2 \times 10^{6}$~s$^{-1}$ and $2 \times 10^{14}$~s$^{-1}$. 
However, the laws of variation of the characteristic times as a function of $\nu$ will allow us to extrapolate them to smaller tunneling rates $\nu$ which often characterize telegraphic noises \cite{Paladino14}.

\begin{table}  
  \caption{Charge traps considered in this paper. Position: The coordinates $x$, $y$ and $z$ are defined with respect to the axes specified in Figs.~\ref{3Ddevice} and \ref{fig_wf_pot}. Characteristics deduced from the perturbation matrix: angular frequency $\omega_{th}$ [Eq.~(\ref{eqn_omega})] and modulus $|u_{\uparrow\downarrow}|$ of the non-diagonal matrix element.}
  \label{table_param}
  \begin{tabular}{|c|ccc|c|c|c|}
    \hline
    Trap & $x$ & $y$ & $z$ & Gate & $\omega_{th}$ & $| u_{\uparrow\downarrow} |$ \\
     & & (nm) &  & & (s$^{-1}$) & ($\mu$eV) \\
    \hline
    Trap 1 & 8.4 & 0.0 & 0.0 & Central & $1.063 \times 10^{9}$ & 1.4594 \\
    Trap 2 & 8.4 & 4.0 & 14.0 & Central & $5.469 \times 10^{8}$ & 0.4381 \\
    Trap 3 & 8.4 & 0.0 & 46.0 & Secondary & $3.039 \times 10^{7}$ & 0.0248 \\
    \hline
  \end{tabular}
\end{table}

We have considered three positions for the trap (Table~\ref{table_param}). 
Trap 1 and Trap 2 are under the central gate, Trap 3 is under the secondary gate. Trap 1 is the closest to the hole quantum dot. 
It therefore induces the strongest perturbation potential on the hole.
In contrast, Trap 3 induces the lowest perturbation. 

\section{Results and discussion}

\subsection{Quantum dot energy levels and hole state dynamics}

\begin{figure*}%[!h]
\centering
\includegraphics[width=1.7\columnwidth]{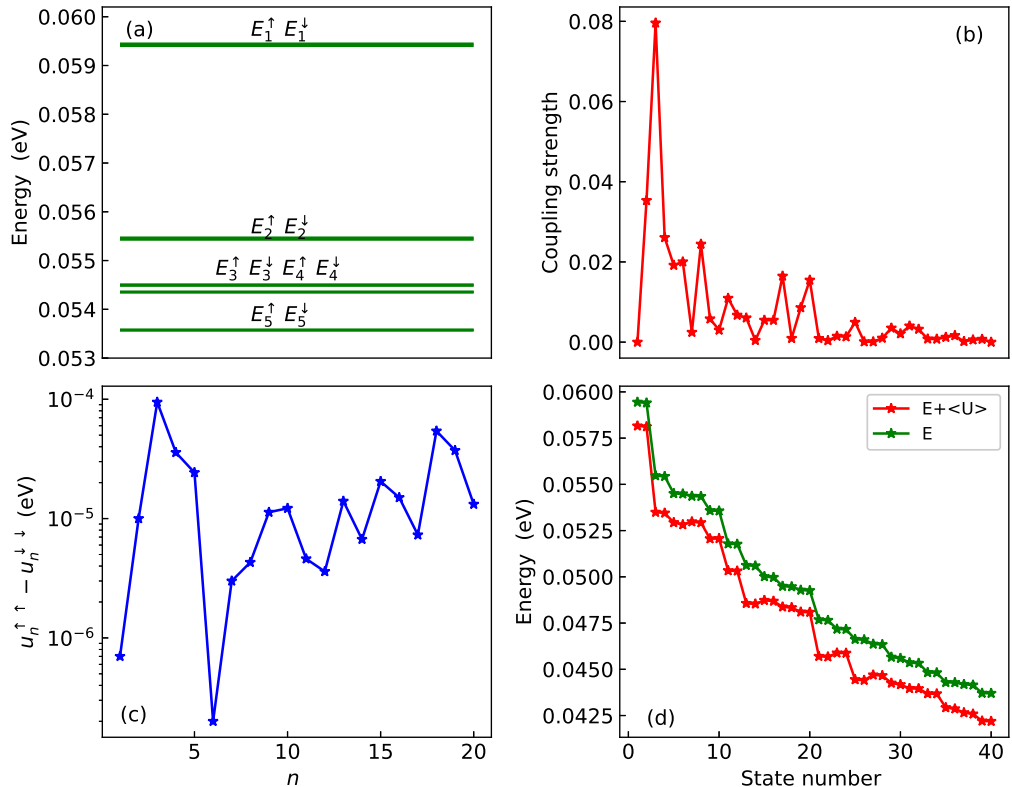}
\caption{(a) Highest electronic energy levels calculated for the hole qubit. 
(b) Coupling strength defined as the ratio $\left|\langle \varphi_{1}^{\uparrow} | U | \varphi_{n}^{\uparrow\downarrow} \rangle \right|/|E_{1}^{\uparrow} - E_{n}^{\uparrow\downarrow}|$. 
(c) $\delta_{n} = \langle \varphi_{n}^{\uparrow} | U | \varphi_{n}^{\uparrow} \rangle - \langle \varphi_{n}^{\downarrow} | U | \varphi_{n}^{\downarrow} \rangle =  u_{n}^{\uparrow\uparrow} - u_{n}^{\downarrow\downarrow}$ versus $n$.
(d) Unperturbed level energies $E_{n}^{\uparrow\downarrow}$ (green) and perturbed level energies $E_{n}^{\uparrow\downarrow}+\langle \varphi_{n}^{\uparrow\downarrow} | U | \varphi_{n}^{\uparrow\downarrow} \rangle$ (red) presented according to the state number defined as $2n-1$ for $|\varphi_{n}^{\uparrow}\rangle$ states and $2n$ for $|\varphi_{n}^{\downarrow}\rangle$ states. (b-d) All results are for Trap 1.
(b) and (d) share the same horizontal axis.}
\label{fig_levels_couplings}
\end{figure*}

\begin{figure}%[!h]
\centering
\includegraphics[width=0.90\columnwidth]{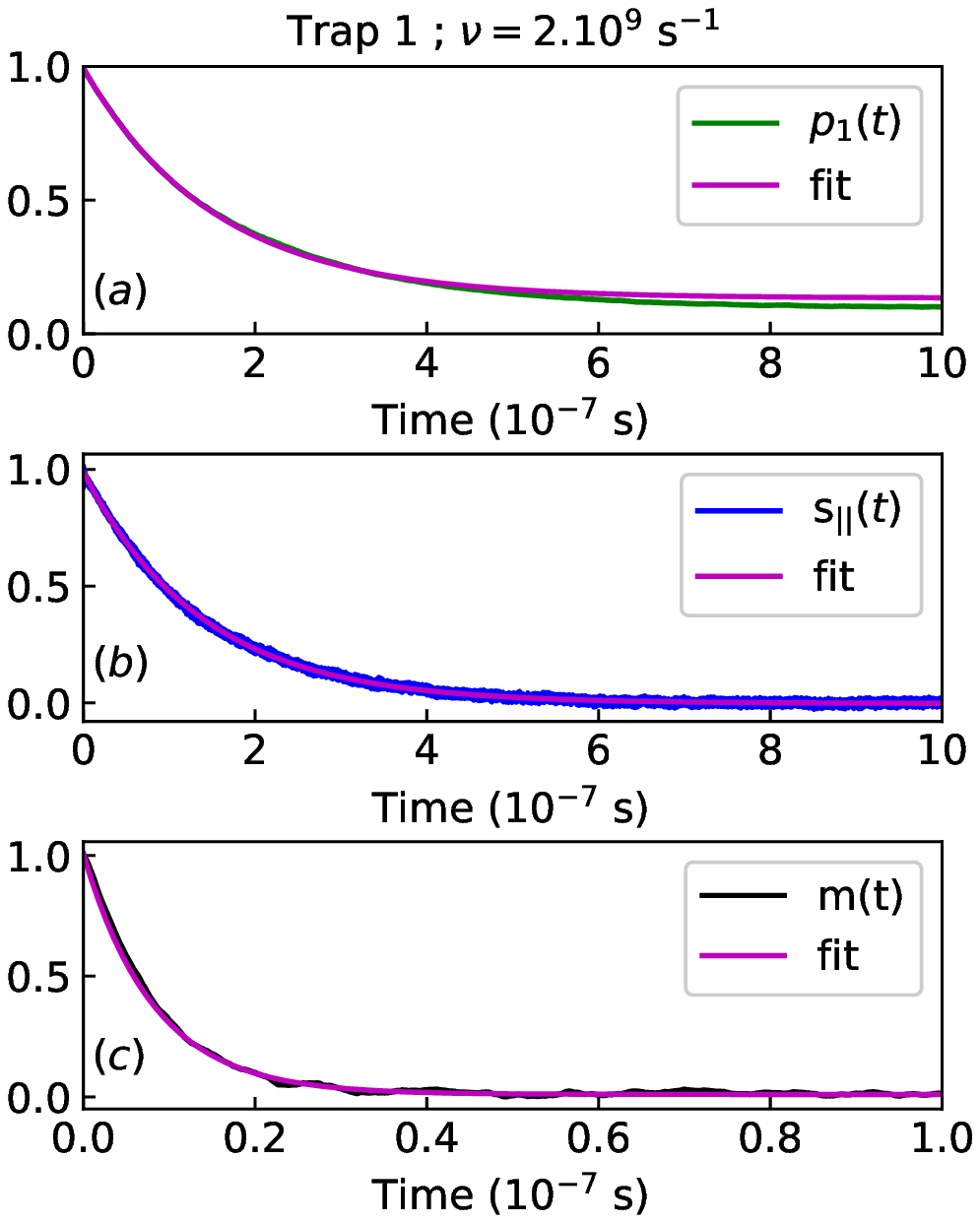}
\caption{Evolution of $p_{1}(t)$ (a), $\sigma_{\parallel}(t)$ (b) and $m(t)$ (c) for Trap 1 and $\nu=2 \times 10^{9}$~s$^{-1}$, calculated in the multi-level model ($N=20$). The curves in pink are fits by an exponential law. For $p_{1}(t)$, the fit is made in the range $0.5\le p_1\le 1$ only.}
\label{fig_fit_lifetimes}
\end{figure}

The 10 highest electronic energy levels calculated for the qubit are shown in Fig.~\ref{fig_levels_couplings}a. The level defining the fundamental hole state, the highest in electron energy, is twice degenerate in the absence of a magnetic field. This level is relatively detached from the others, the Zeeman splitting is not visible at the scale of the figure. Therefore the electrostatically induced confinement by the central gate defines a two-state system.

Nevertheless, Fig.~\ref{fig_fit_lifetimes}a shows the evolution of $p_{1}(t)$, the average total weight of the hole wave function on the $\varphi_{1}^{\sigma}$ qubit states as a function of time, for a charge fluctuating between the central gate and Trap 1 at average frequency $\nu=2 \times 10^{9}$~s$^{-1}$. $p_{1}(t)$ decreases as a function of time, from which we can deduce $T_{1}'$ as described in Appendix~\ref{appendix_cal_t1}. This results from the coupling of the state $\varphi_{1}^{\uparrow}$ with the other states $\varphi_{n}^{\uparrow\downarrow}$ of the system, due to the presence of a charge on Trap 1. Interestingly, this coupling is much larger with some states outside the doublet $(n>1)$ than with $\varphi_{1}^{\downarrow}$.
This is demonstrated in Fig.~\ref{fig_levels_couplings}b, which shows the coupling strength defined as the ratio between a matrix element of $U$ and the energy splitting between the corresponding two states. This coupling strength is important between states of the same doublet (because the denominator is small), but it remains of the same order of magnitude with a large number of multiplets much further away in energy.

Figures~\ref{fig_fit_lifetimes}b and \ref{fig_fit_lifetimes}c show the decay of $\sigma_{\parallel}(t)$ and $m(t)$ obtained under the same noise conditions. We deduce the characteristic times $T_1$ and $T_2$ by fitting with an exponential.

Figure~\ref{fig_levels_couplings}d shows that diagonal terms of the perturbation $u_{n}^{\uparrow\uparrow}=\langle \varphi_{n}^{\uparrow} | U | \varphi_{n}^{\uparrow} \rangle$ or $u_{n}^{\downarrow\downarrow}=\langle \varphi_{n}^{\downarrow} | U | \varphi_{n}^{\downarrow} \rangle$ are relatively independent of the spin orientation in each doublet $n$, their main effect being a global shift in energy of the electronic levels. The difference $\delta_{n} = u_{n}^{\uparrow\uparrow}  - u_{n}^{\downarrow\downarrow}$, which for $n=1$ will determine the main dephasing effect (see below), is small and is strongly dependent on $n$ (Fig.~\ref{fig_levels_couplings}c). 
It is shown in Sect.~II of the Supplemental Material \cite{supplemental} that $\delta_{n}$ is zero for $B=0$ and is given in perturbation theory by a sum of terms scaling as  $|E_{n}-E_{m}|^{-1}$ with $m\neq n$.
This explains why $\delta_{1}= u_{1}^{\uparrow\uparrow}  - u_{1}^{\downarrow\downarrow} = u_{\uparrow\uparrow}  - u_{\downarrow\downarrow}$ is small because the fundamental level $(n=1)$ is strongly detached from the others in energy.

\subsection{Characteristic times for the two-level system}

Even if our numerical simulations show the non negligible role of high energy hole states, it is useful to consider the two-level system $(N=1)$.
Relaxation $(T_{1})$ and dephasing $(T_{2})$  times calculated for the Trap 1 (Table~\ref{table_param}) in the two-level model are therefore presented in Fig.~\ref{fig_T1_T2_two_level}.

\subsubsection{Analytical model: Dephasing time}

It is interesting to relate the variations of $T_2(\nu)$ to those of $T_{2}^{*}(\nu)$, the dephasing time obtained in the so-called pure dephasing model, i.e. when the matrix $U$ [Eq.~(\ref{huzero})] is purely diagonal, or more generally when $|u_{\uparrow\uparrow}-u_{\downarrow\downarrow}| \gg |u_{\uparrow\downarrow}|$. In this case, the (threshold) angular frequency can be written as (Sect.~III of the Supplemental Material \cite{supplemental})

\begin{equation}
\omega_{th} \approx |u_{\uparrow\uparrow} - u_{\downarrow\downarrow}|/\hbar
\label{eqn_omega}
\end{equation}

\noindent in which $|u_{\uparrow\uparrow} - u_{\downarrow\downarrow}|$ ($\delta_{1}$ in Fig.~\ref{fig_levels_couplings}c) represents the variation of the energy splitting between the two states when the fluctuator switches.
In the high frequency limit $(\nu \gg \omega_{th})$, the phase undergoes many random changes $\delta\phi(t)$ over a time interval of the order of $2\pi/\omega_{th}$, so that $\delta\phi(t)$ can be viewed as a continuous random variable characterized by a Gaussian probability distribution. In this Gaussian limit, the dephasing time is given by $T_{2}^{*} = 4\nu/\omega_{th}^{2}$ \cite{Abel08,abelthese,Bergli09}. The linear dependence on the frequency $\nu$ reflects the fact that the two-level system becomes more and more insensitive to the random perturbation as this one varies more and more rapidly. In this case, the splitting between the two levels is self-averaged to a certain value, the width being $\propto 1/T_{2}^{*}$ \cite{Bergli09}.

In the opposite limit of low frequencies $(\nu \ll \omega_{th})$, the Gaussian approximation is no longer valid and the dephasing time is then given by $T_{2}^{*} = 2/\nu = 1/\nu_{cl}$ \cite{Abel08,abelthese,Bergli09} (see also Sect.~III of the Supplemental Material \cite{supplemental}). In this regime, the coherence is simply determined by the probability $\exp (-t/T_{2}^{*})$ that the qubit has not suffered any switch of the fluctuator over a time $t$. In other words, the coherence is lost from the moment when the fluctuator has changed state.

Extending the analysis to all frequencies, an expression for $T_{2}^{*}$ was derived in the pure dephasing regime  \cite{Abel08,abelthese}:

\begin{equation}
T_{2}^{*} = \bigg\{ 
\begin{array}{c}
\frac{2} {\nu - \sqrt{\nu^{2} - \omega_{th}^{2}}} \textrm{~for~} \nu \ge \omega_{th} \\ 
2/\nu \textrm{~for~} \nu < \omega_{th}
\end{array}
\label{eq_anal_T2}
\end{equation}

\noindent which shows that the angular frequency $\omega_{th}=|\Omega-\Omega'|$ is therefore a threshold between two distinct regimes.

In the general case, beyond the pure dephasing model, the dephasing time becomes \cite{Paladino14}

\begin{equation}
T_{2} = \left(\frac{1}{T_{2}^{*}} + \frac{1}{2T_{1}} \right)^{-1}. 
\label{eq_T2_T1_T2*}
\end{equation}

\subsubsection{Analytical model: Relaxation time}
The spin relaxation is induced by the non-diagonal terms of the $U$ matrix. Using various approaches such as the Bloch-Redfield theory \cite{Paladino14}, the relaxation rate is determined by the noise spectrum at frequency $\Omega$. In the case of telegraphic noise, this gives the following expression for $T_1$,

\begin{equation}
T_{1} = \frac{\hbar^{2}}{|u_{\uparrow\downarrow}|^{2}} \frac{\nu^{2} + \Omega^{2}}{\nu} 
\label{eq_anal_T1}
\end{equation}

\noindent which reflects a resonance effect. Relaxation is most effective when the fluctuator frequency coincides with the Larmor angular frequency.
At low tunneling rate $\nu \ll \omega_{th} \ll \Omega$, $T_1$ also varies as $1/\nu$, like $T_{2}^{*}$ but with a prefactor $\left( \hbar \Omega/|u_{\uparrow\downarrow}| \right)^{2}$ instead of $2$ (another derivation of this expression is presented in Sect.~V of the Supplemental Material \cite{supplemental}). 
Table~\ref{table_param} shows that $|u_{\uparrow\downarrow}| \ll \hbar \Omega$ for the traps considered here,  therefore $T_{1} \gg T_{2}^{*}$ and $T_{2} \approx T_{2}^{*}$ from Eq.~(\ref{eq_T2_T1_T2*}).

\subsubsection{Discussion of numerical results in comparison with analytical laws}

\begin{figure}%[!h]
\centering
\includegraphics[width=0.8\columnwidth]{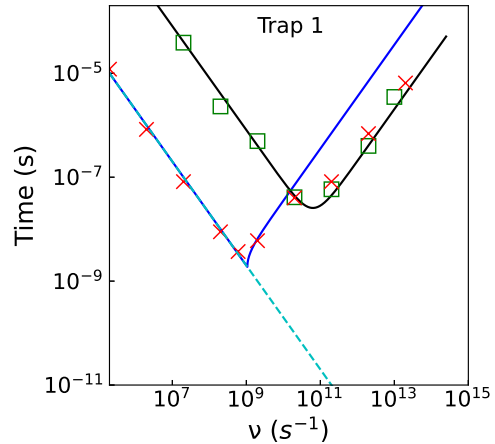}
\caption{Characteristic lifetimes $T_1$ (green squares) and $T_2$ (red crosses) versus tunneling rate $\nu$ calculated in the two-level model for Trap 1. Solid lines depict the analytical expressions for $T_1$ (black) and $T_{2}^{*}$ (light blue), as given by Eq.~(\ref{eq_anal_T1}) and Eq.~(\ref{eq_anal_T2}), respectively, using $\omega_{th}$ and $|u_{\uparrow\downarrow}|$ of Table~\ref{table_param}. The dashed turquoise line shows a time varying as $2/\nu$. At $\nu \gg \Omega$, $T_{2} \approx 2T_{1}$.}
\label{fig_T1_T2_two_level}
\end{figure}

Figure~\ref{fig_T1_T2_two_level} shows that the calculated values of $T_1$ follow Eq.~(\ref{eq_anal_T1}) which translates the resonance effect between the quantum oscillator formed by the two-level system and the classical fluctuator. 

The dephasing time $T_2$ deduced from the time simulations is also in excellent agreement with the analytical expression for $T_{2}^{*}$ given by Eq.~(\ref{eq_anal_T2}), except at high frequency where $T_1$ becomes smaller than $T_{2}^{*}$ and therefore $T_{2} \approx 2T_{1}$ [Eq.~(\ref{eq_T2_T1_T2*})].

\subsection{Characteristic times for the multi-level system}

\begin{figure}%[!h]
\centering
\includegraphics[width=0.8\columnwidth]{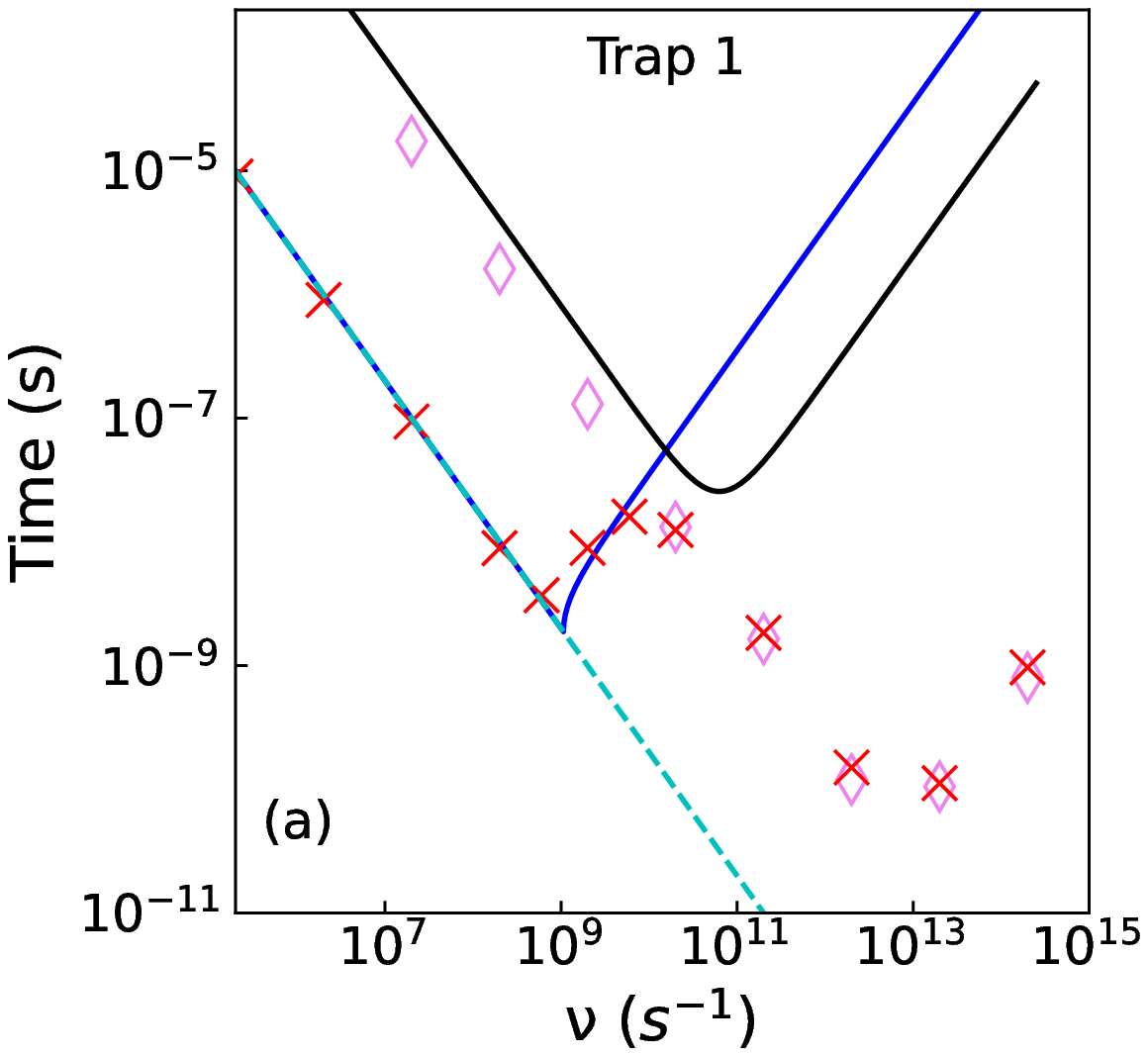}
\includegraphics[width=0.8\columnwidth]{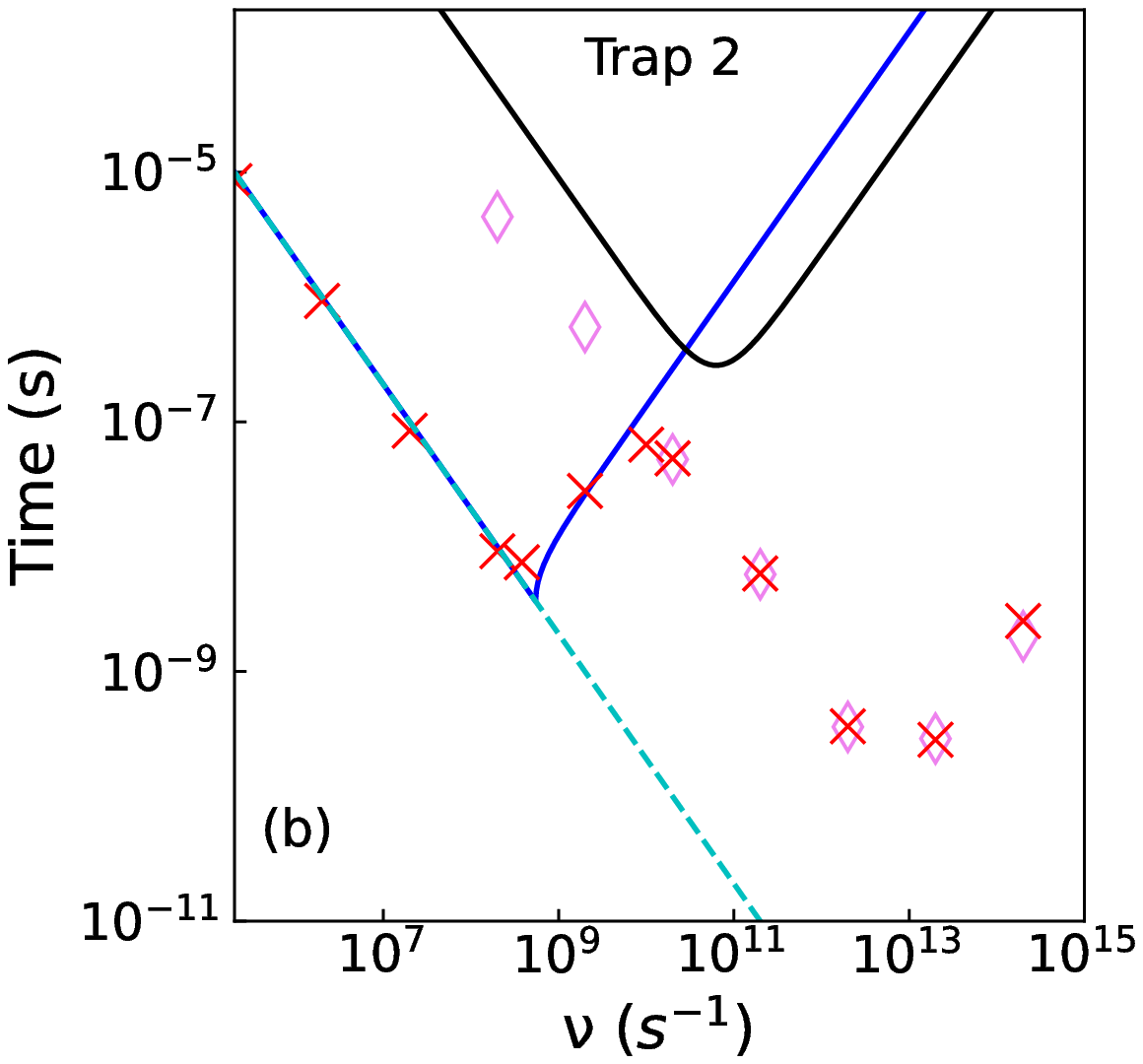}
\includegraphics[width=0.8\columnwidth]{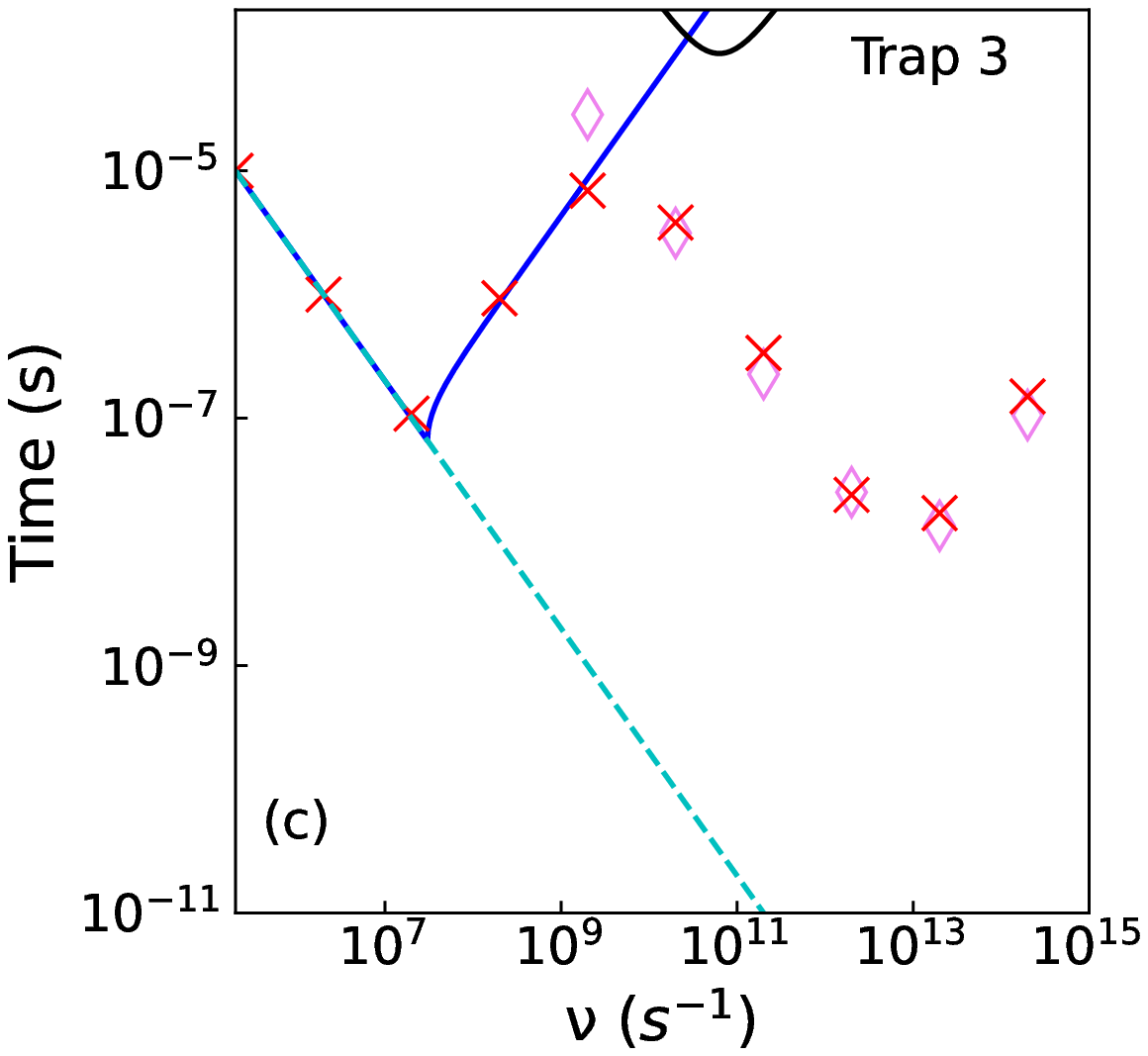}
\caption{(a) Characteristic lifetimes $T_{1}'$ (magenta diamonds) and $T_2$ (red crosses) versus tunneling rate $\nu$ calculated in the multi-level model $(N=10)$ for Trap 1. Solid lines depict the analytical expressions for $T_1$ (black) and $T_{2}^{*}$ (blue) of the two-level model, as given by Eq.~(\ref{eq_anal_T1}) and Eq.~(\ref{eq_anal_T2}), respectively, using $\omega_{th}$ and $|u_{\uparrow\downarrow}|$ of Table~\ref{table_param}. (b) Same for Trap 2. (c) Same for Trap 3. (a-c) The dashed turquoise line shows a time varying as $2/\nu$.}
\label{fig_T1_T2_multi_level}
\end{figure}

The calculated characteristic times $T_2$ and $T_{1}'$ for the three traps in the multi-level model are shown in Fig.~\ref{fig_T1_T2_multi_level}. 

\subsubsection{Relaxation time for Trap 1}

The relaxation time $T_{1}'$ is much smaller than $T_1$ obtained in the two-level model, in particular at high frequency ($>10^{10}$~s$^{-1}$) where $T_{1}'$ continues to decrease with $\nu$ to reach a minimum for $\nu$ near $10^{13}$~s$^{-1}$.
This is due to the coupling with higher energy hole levels, outside the doublet.
For frequencies below $\approx 10^{12}$~s$^{-1}$, $T_{1}'$ varies approximately as $1/\nu$, as given by Eq.~(\ref{eq_anal_T1}) for $T_1$ for $\nu \ll \Omega$ but with a smaller prefactor.

The fact that $T_{1}'$ is found much smaller than $T_{1}$ given by Eq.~(\ref{eq_anal_T1}) means that the two-level model is not valid for the description of the spin relaxation, the latter being strongly influenced by the coupling to higher energy hole levels.   

\subsubsection{Dephasing time for Trap 1}

Remarkably, the values of $T_2$ coincide with those obtained in the two-level model for $\nu$ less than or just above $\omega_{th}$ given in Table~\ref{table_param}. 
In this case, the two-level model is perfectly justified.
For higher tunneling rates, the values of $T_2$ approximately follow those of $T_{1}'$. This behavior indicates that the dephasing is impacted by the other decoherence phenomena, which is expected because they are fundamentally intertwined since they all result from the same electrical disturbance, i.e., diagonal and non-diagonal terms are present at the same time in the matrix of $U$.

\subsubsection{Results for Trap 2 and 3}

Figure~\ref{fig_T1_T2_multi_level}b shows the same behavior for Trap 2 located at a larger distance from the center of the qubit state. Consequently, the characteristic times $T_1$ (for the two-level model) and $T_{1}'$ have higher values than for Trap 1, since the perturbation induced by the fluctuator is less strong. For the same reason, $\omega_{th}$ shifts to a lower frequency. 
This behavior is even more visible in the case of Trap 3 located under the secondary gate (Fig.~\ref{fig_T1_T2_multi_level}c). $\omega_{th}$ is pushed to even lower frequencies (Table~\ref{table_param}). 
The comparison between Fig.~\ref{fig_T1_T2_multi_level}a, Fig.~\ref{fig_T1_T2_multi_level}b and Fig.~\ref{fig_T1_T2_multi_level}c shows that the two-level model for $T_1$ becomes even less valid when the trap moves away from the qubit, when $T_1$ becomes very long. 
Indeed, as the distance between the trap and the qubit increases, the coupling terms all decrease, but those with the higher energy states decrease less rapidly than those within the doublet of states (Sect.~VI of the Supplemental Material \cite{supplemental}). 
This can be explained by the larger spatial extension of higher energy hole states.

A remarkable result of Fig.~\ref{fig_T1_T2_multi_level} is that, for $\nu < \omega_{th}$, $T_{2} \approx T_{2}^{*}$ is given by $2/\nu = 1/\nu_{cl}$, whatever the position of the trap, only the value of $\omega_{th}$ changes between the different cases.
This is the likely regime for a qubit in cryogenic conditions for which the tunneling rates are normally low \cite{Bergli09,Clerk10,Paladino14}. 
This means that the coherence of the qubit is entirely and solely determined by the average time between two changes of state of the fluctuator.
The qubit remains coherent as long as the fluctuator has not changed its state.

\section{Influence of the back gate bias and the magnetic field orientation}

In a recent theoretical work including numerical simulations on the same hole qubit as the one studied here \cite{Venitucci18}, the manipulation of the hole spin by a radio-frequency electrical excitation applied on the central gate has been modeled.  It has been shown that the Rabi frequency depends in a complex way on the orientation of the magnetic field and the back-gate potential $V_{BG}$. The latter allows one to control the shape and the symmetry of the wave function of the hole on which depends the effective spin-orbit coupling felt and consequently the $g$ tensor defining the response of its spin to the magnetic field. 

In fact, all these quantities depend essentially on the component of the internal electric field along $y$, which is controlled by the imbalance between front and back gate voltages \cite{Venitucci18}. This component is non-zero because of the asymmetry of the structure (Fig.~\ref{3Ddevice}). The component of the field along $x$, although dominant in intensity, plays a much more minor role due to the strong vertical confinement of the hole states. The $y$ component of the field influences the wave function of the ground state of the hole, not only its position along $y$ but also the respective weight of heavy and light hole components, which determines the effective spin-orbit coupling applying to the hole  (Sect.~IV of the Supplemental Material \cite{supplemental}). 

Interestingly, it was found in Ref.~\cite{Venitucci18} that, for $V_{BG}=-0.15$ V, the qubit is placed in a configuration where the spin is largely insensitive to radio-frequency excitation on the central gate, the Rabi frequency showing a sharp minimum. In this case, the wave function of the hole is centered in the cross section of the nanowire and presents an approximate inversion symmetry which tends to reduce the action of the spin-orbit coupling on the hole. At this voltage, the influence of the Johnson-Nyquist noise is minimized, as well as the coupling to phonons but to a lesser extent \cite{Li20}.

In this context, we consider the effect of $V_{BG}$ and the magnetic field orientation on the dephasing time $T_2$. As discussed earlier, the evolution of $T_2$ as a function of $\nu$ is defined by an angular frequency $\omega_{th}$ which delineates the low and high frequency regimes of the spin response to the fluctuator. 
It is therefore interesting to look for situations where $u_{\uparrow\uparrow}-u_{\downarrow\downarrow}=0$ and therefore $T_{2}(\omega_{th}) = 2/\omega_{th} = 2\hbar/|u_{\uparrow\uparrow}-u_{\downarrow\downarrow}|$ should diverge.

\begin{figure}%[!h]
\centering
\includegraphics[width=0.8\columnwidth]{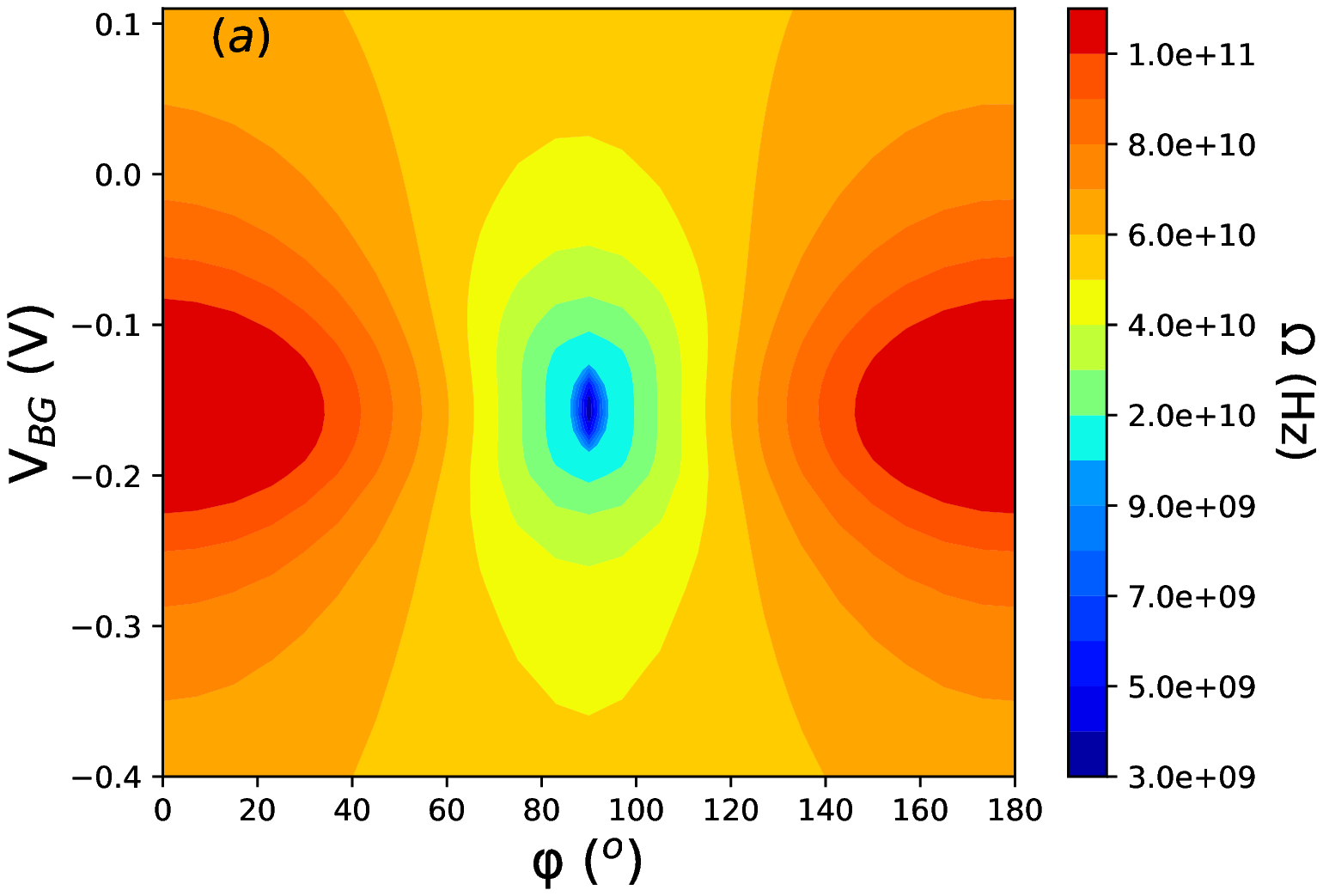}
\includegraphics[width=0.8\columnwidth]{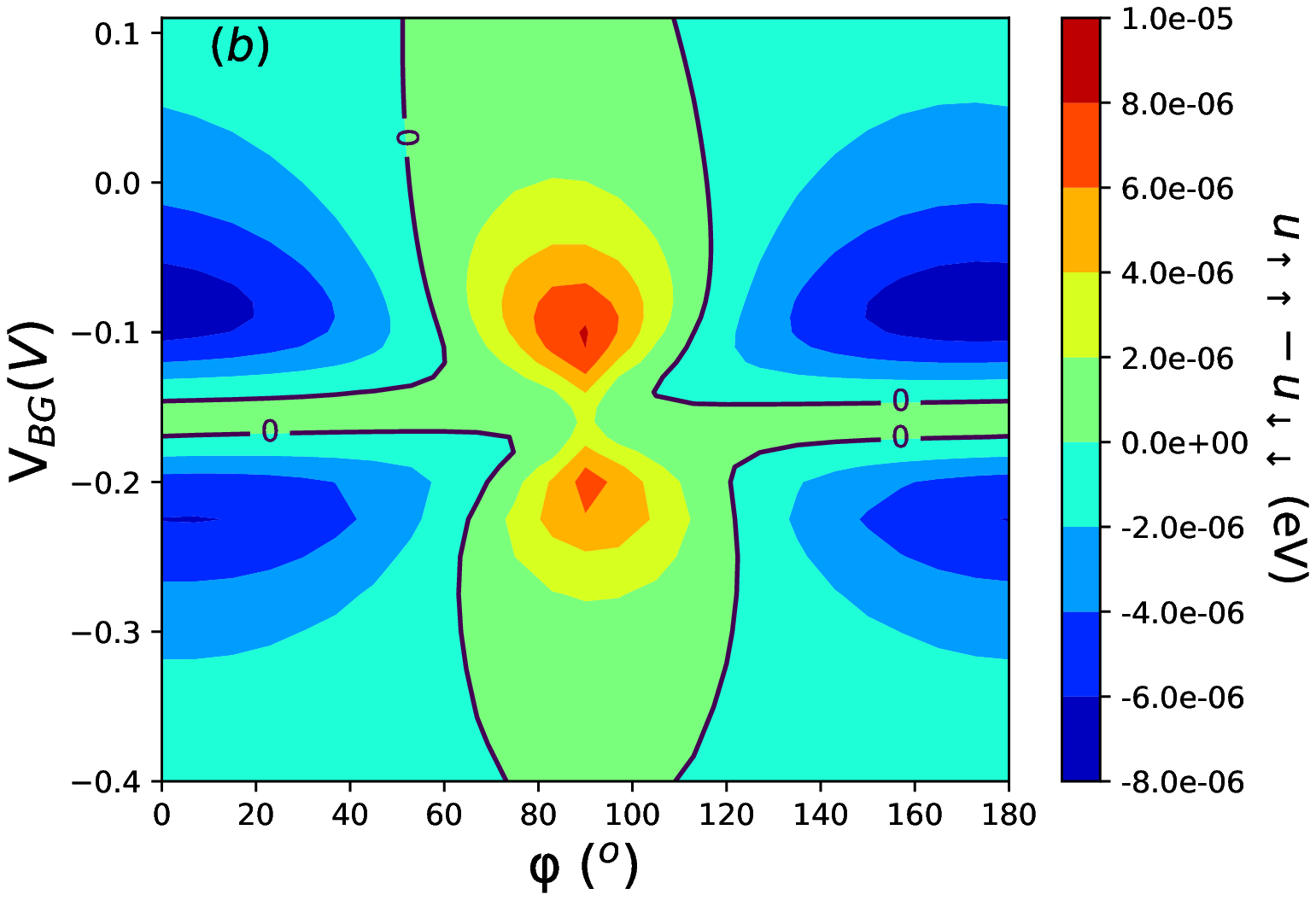}
\includegraphics[width=0.8\columnwidth]{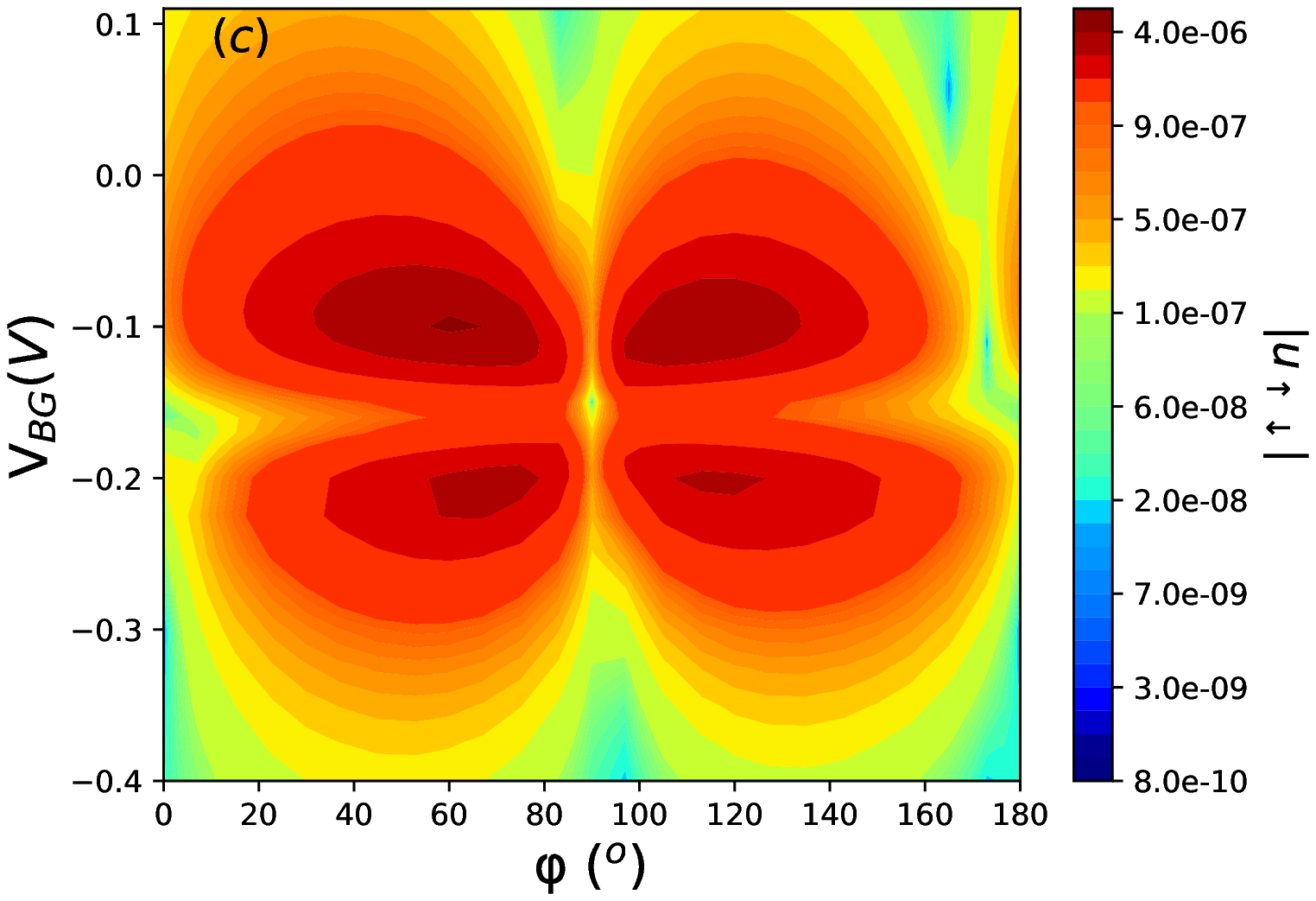}
\caption{2D plots of (a) the Larmor frequency $\Omega$, (b) $u_{\uparrow\uparrow}-u_{\downarrow\downarrow}$ and (c) $|u_{\uparrow\downarrow}|$ for Trap 1 versus back gate bias $V_{BG}$ and angle $\varphi$ of $B$ in the $xy$ plane ($\theta=90^{\circ}$). The 2D plots are made on a discrete grid of $25 \times 40$ points. (b) The contours corresponding to $u_{\uparrow\uparrow}-u_{\downarrow\downarrow}=0$ are indicated by black lines. }
\label{fig_Omega_u_Vbg_phi_Trap1}
\end{figure}

Figure~\ref{fig_Omega_u_Vbg_phi_Trap1} presents the variations of $\Omega$ and $u_{\uparrow\uparrow}-u_{\downarrow\downarrow}$ as a function of $V_{BG}$ and $\phi$, for $\theta=90^{\circ}$, i.e. for a magnetic field in the plane perpendicular to the axis of the nanowire, the field strength remaining fixed at its initial value.
Figure~\ref{fig_Omega_u_Vbg_phi_Trap1}b shows that the 2D map $u_{\uparrow\uparrow}-u_{\downarrow\downarrow} = f(V_{BG},\varphi)$ is divided into regions of positive or negative values, which means that $u_{\uparrow\uparrow}-u_{\downarrow\downarrow}=0$ at the boundaries.
Consequently, this result suggests the existence of "sweet" lines \cite{Piot22,Michal22} along which $T_{2}(\omega_{th})$ should become very long (but not infinite for reasons discussed in Sect.~\ref{sect_disc_T2}).

Zeros of $u_{\uparrow\uparrow}-u_{\downarrow\downarrow}$ are present in particular along a horizontal line $V_{BG} \approx -0.15$~V, in the same configuration where the spin becomes relatively insensitive to electrical noise on the central gate \cite{Venitucci18}.
Remarkably, there are also two (almost) straight vertical lines for which $u_{\uparrow\uparrow}-u_{\downarrow\downarrow} \approx 0$, at $\varphi \approx 55^{\circ}$ and $\varphi \approx 125^{\circ}$.
These can be explained as follows.

As demonstrated in Ref.~\cite{Venitucci18} and in Sect.~IV of the Supplemental Material \cite{supplemental}, the effective Zeeman Hamiltonian of the system can be written in the $g$-matrix formalism. For $\vec{B}$ in the $xy$ plane, the Zeeman splitting is

\begin{equation}
\hbar \Omega = \mu_{B} B \sqrt{g_{x}^{2} \cos^{2}\varphi + g_{y}^{2} \sin^{2}\varphi}.
\label{eqn_Zeeman_splitting}
\end{equation} 

It is shown in Ref.~\cite{Piot22} that the respective weight of the hole wave function on the heavy and light hole states determines the relative magnitude of the factors $g_x$ and $g_y$. It follows that

\begin{equation}
\frac{\partial g_{x}}{\partial V} \approx - \frac{\partial g_{y}}{\partial V}
\label{eq_deriv_g}
\end{equation} 

\noindent where $V$ can be any potential whose main effect on the $g$ factors comes from the variation of the electric field along $y$. Our calculations show that this is the case for $V_{BG}$ or for the potential induced by the fluctuating charge.

Under these conditions, it is interesting to consider situations where the Zeeman splitting [Eq.~(\ref{eqn_Zeeman_splitting})] becomes relatively independent of the potential $V_{BG}$.
Using Eq.~(\ref{eq_deriv_g}) with $V=V_{BG}$, we deduce in Sect.~IV of the Supplemental Material \cite{supplemental} that $(\partial \,\hbar \Omega) / (\partial \,V_{BG}) = 0$ for 

\begin{equation}
\varphi \approx \frac{\pi}{2} \pm \arctan \sqrt{\frac{g_x}{g_y}}. \label{eqn_vertical_lines_2}
\end{equation}

Therefore, the compensation between $\partial g_{x}/\partial V$ and $\partial g_{y}/\partial V$ [Eq.~(\ref{eq_deriv_g})] leading to Eq.~(\ref{eqn_vertical_lines_2}) explains the straight vertical contour lines in the 2D plot of the Larmor frequency at $\varphi \approx 90 \pm 34^{\circ}$ for $g_{x}/g_{y} \approx 2/3$ (Fig.~\ref{fig_Omega_u_Vbg_phi_Trap1}a). The existence of these sweet lines has been revealed experimentally in Ref.~\cite{Piot22}, where a strong enhancement of the coherence times was measured when the Larmor frequency $\hbar \Omega$ is least dependent on the gate voltages.

The $g$-matrix model also allows us to derive analytical expressions for the diagonal matrix elements $u_{\uparrow\uparrow}$ and $u_{\downarrow\downarrow}$ of $\delta H$ describing the perturbation brought by the fluctuating charge (Sect.~IV of the Supplemental Material \cite{supplemental}). 
Remarkably, we find in this simplified model that $u_{\uparrow\uparrow}-u_{\downarrow\downarrow}$ cancels out when Eq.~(\ref{eq_deriv_g}) and Eq.~(\ref{eqn_vertical_lines_2}) are verified exactly, $V$ representing the potential induced by the charge.

Additional 2D plots of $\Omega$ and $u_{\uparrow\uparrow}-u_{\downarrow\downarrow}$ are presented in Sect.~VII of the Supplemental Material \cite{supplemental}, versus $\theta$ and $\varphi$, for the three traps. "Sweet" lines are clearly visible on these figures for $\varphi$ approximately given by Eq.~(\ref{eqn_vertical_lines_2}).
We conclude that their existence is relatively robust to the position of the trap \cite{Piot22}.

\section{Discussion of  the results}

\subsection{Discussion of $T_{2}$}
\label{sect_disc_T2}

The existence of "sweet" spots where the dephasing of the spin precession is strongly reduced is now much discussed in the literature \cite{Venitucci18,Venitucci19,Benito19,Bosco21,Wang21,Adelsberger22,Piot22,Michal22}.
The results presented in the previous section could suggest the existence of "sweet" lines [in the $(V_{BG},\varphi)$ or $(\theta,\varphi)$ operating spaces] for the noise induced by a single charge fluctuator, but this conclusion should be immediately relativized, for two reasons. 

First, for a "slow" fluctuator ($\nu$ smaller than $\omega_{th}$), $T_{2}(\nu)$ is given by $2/\nu$, independently of the field or potential conditions.
Second, even if we place the system in a situation where $u_{\uparrow\uparrow}-u_{\downarrow\downarrow}=0$, this does not mean that $\omega_{th} \to 0$ and $T_{2} \to \infty$ because the non-diagonal term $u_{\uparrow\downarrow}$ is not zero (and is even usually maximal \cite{Michal22}).
This is clearly visible by comparing, in Fig.~\ref{fig_Omega_u_Vbg_phi_Trap1}b and Fig.~\ref{fig_Omega_u_Vbg_phi_Trap1}c, $u_{\uparrow\downarrow} \neq 0$ along the lines where $u_{\uparrow\uparrow}-u_{\downarrow\downarrow}=0$. 
We show in Sect.~III of the Supplemental Material \cite{supplemental} that the angular frequency $\omega_{th}=|\Omega-\Omega'|$ is given in the case $|u_{\uparrow\uparrow} - u_{\downarrow\downarrow}| \ll |u_{\uparrow\downarrow}|$ by

\begin{equation}
\omega_{th} \approx \frac{2|u_{\uparrow\downarrow}|^{2}}{\hbar^{2}\Omega}.
\label{eq_wth_no_diag}
\end{equation}

It is interesting to note that $\omega_{th}$ (and therefore $T_{2}(\omega_{th}) = 2/\omega_{th}$) given in Eq.~(\ref{eq_wth_no_diag}) is not of same order in $U$ as in Eq.~(\ref{eqn_omega}).
The physics discussed here only reveals itself when $u_{\uparrow\uparrow}-u_{\downarrow\downarrow} \approx 0$.

\begin{figure}%[!h]
\centering
\includegraphics[width=0.8\columnwidth]{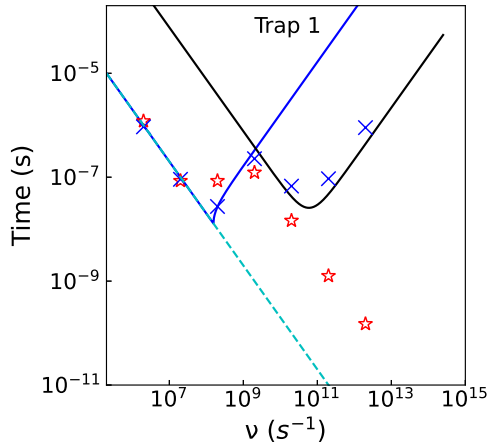}
\caption{Characteristic lifetimes $T_2$  versus tunneling rate $\nu$ calculated in the two-level model (blue crosses) and multi-level $(N=10)$ model (red stars) for Trap 1, for $V_{BG}=0$~V, $\varphi=52.6^{\circ}$, $\theta=90^{\circ}$ and $\hbar\Omega = 3.88 \times 10^{-5}$~eV, in a situation where $u_{\uparrow\uparrow} - u_{\downarrow\downarrow} \approx 0$~eV and $|u_{\uparrow\downarrow}|=1.41 \times 10^{-6}$~eV. The black solid line depicts the analytical expression for $T_1$ of the two-level model, as given by Eq.~(\ref{eq_anal_T1}). The dashed turquoise line shows a time varying as $2/\nu$.}
\label{fig_T2_sweet_spot}
\end{figure}

Figure~\ref{fig_T2_sweet_spot} shows the evolution of $T_2$ as a function of $\nu$ when the system is placed at a specific point on a "sweet" line of Fig.~\ref{fig_Omega_u_Vbg_phi_Trap1}b. In the chosen case, Eq.~(\ref{eq_wth_no_diag}) gives $\omega_{th} = 1.6 \times 10^{8}$~s$^{-1}$ for the threshold angular frequency.
This value is about one order of magnitude smaller than in Table~\ref{table_param}, showing the interest to be along a "sweet" line.
For $\nu \ll \omega_{th}$, we again find that $T_2$ behaves like $2/\nu$ for reasons discussed in Sect.~III of the Supplemental Material \cite{supplemental}. 
Beyond that, in the two-level model, $T_2$ increases to reach values close to $2T_1$, the dephasing process is limited by the spin relaxation. 
In the multi-level model, the high-frequency regime is once again influenced by coupling to higher energy states.
We thus find a behavior identical to the one obtained in the case where $u_{\uparrow\uparrow}-u_{\downarrow\downarrow} \neq 0$, but the value of $\omega_{th}$ is here determined by the non-diagonal element of the perturbation Hamiltonian.
However, it is interesting to note that, even if the non-diagonal term $u_{\uparrow\downarrow}$ is responsible for the spin relaxation phenomenon, the value of $T_2$ at frequency $\omega_{th}$ [Eq.~\ref{eq_wth_no_diag}] is much lower than $T_1$ [Eq.~\ref{eq_anal_T1}] at this same frequency:

\begin{equation}
T_{1}(\omega_{th}) = T_{2}(\omega_{th}) \frac{\hbar^{2}\Omega^{2}}{2|u_{\uparrow\downarrow}|^{2}} \gg T_{2}(\omega_{th}).
\end{equation}

The combined examination of Fig.~\ref{fig_Omega_u_Vbg_phi_Trap1}a and Fig.~\ref{fig_Omega_u_Vbg_phi_Trap1}b shows that it would be possible to select points along the lines $u_{\uparrow\uparrow}-u_{\downarrow\downarrow} = 0$ where the values of $|u_{\uparrow\downarrow}|$ and thus of $\omega_{th}$ are even smaller, for example by increasing the value of $|V_{BG}|$. In this case, the wave function of the hole is compressed against the side edges of the wire \cite{Piot22}.
This increases confinement, hence splits $E_1$ from the other energy levels, thus reducing the diagonal and non-diagonal coupling terms (Section II.B of the Supplemental Material \cite{supplemental}). An alternative option to increase confinement would be to reduce the channel thickness and gate length, which is a real technological challenge given the already small dimensions of the current devices.

\subsection{Role of the direct Rashba effect}

Several theoretical works have shown that hole bands in Si or Ge/Si nanowires can be characterized by a strong Rashba effect (called "direct") under the action of an electric field \cite{Kloeffel11,Kloeffel18,Luo17}. This raises the question of the importance of this effect on spin decoherence in the qubit studied here. The Rashba interactions couple the spin to the momentum of the particle and can, therefore, bring additional dephasing while the dot is moving at finite speed following a change of state of the fluctuator. This process is naturally included in our time-dependent simulations of the multi-level model that describe the full dynamics of the wave function.
 
In the two-level model, the spin phase drift is induced by the succession of "quasi-static" configurations with different Larmor frequencies $\Omega$ and $\Omega'$. Each configuration is characterized by stationary states given by the diagonalization of the $2 \times 2$ matrices $H_0$ and $H_{0}+U$, respectively. So the phase decoherence results from the "deformation" of the wave function between the two configurations (as a rigid block displacement of the wave function does not change the Larmor frequency). We have seen previously that the two-level and multi-level models predict the same values of $T_2$ for frequencies $\nu$ of fluctuators below a certain threshold, as long as $T_{2} \ll T_{1}'$. This suggests that in this regime the "dynamic" effects and thus the direct Rashba effect do not influence $T_2$. It should also be noted that, still in this regime, we obtain almost unchanged values of $T_2$ when we consider a modified telegraph signal in which the transitions $0 \to 1$ and $1 \to 0$ are no longer instantaneous but occur progressively (linearly) over a duration $\Delta t$ of a few picoseconds, even though this duration is much longer than the tunnel time of the order of femtoseconds which is typically admitted (Appendix~\ref{sect_diabatic}). This demonstrates that the phase decoherence in the low frequency regime does not depend on the dynamics of the transition between the two states.

It cannot be ruled out that the Rashba interactions contribute in the high frequency regime for which two-level and multi-level models do not coincide. In this situation, the wave function is subject to very fast noise, $\nu$ being much larger than the angular frequency $\omega_{th}$. But the influence of the direct Rashba effect may be hidden by the fact that $T_2$ is limited by $T_{1}'$, which highlights the complex dynamic effects due to the coupling with higher energy states.

\subsection{Discussion of $T_{1}$}

Our numerical simulations have shown that the relaxation time $T_1$ obtained in a two-level model is not meaningful because the decoherence induced by the coupling to the highest energy states, outside the doublet, is 
in fact faster with a characteristic time $T_{1}'$. However, this does not mean that the real $T_1$, the one that could be measured, is equal to $T_{1}'$ for the following reasons.

The time $T_{1}'$ reflects the fact that the weight of the hole wave function $\psi(t)$ on the states beyond the fundamental doublet tends to grow with $t$, until a final situation where $\psi(t)$ is statistically distributed on all the states of the considered basis (Section~V of the Supplemental Material \cite{supplemental}). This evolution is very progressive, $T_{1}'$ being very large compared to $2/\nu$, the average period of the telegraph signal. In our simulations, this evolution towards the final state is certain (statistically speaking) because the electronic system composed of the qubit and the fluctuator is not perturbed by any dissipative phenomenon. In fact, as the energy of the electronic system increases on average continuously, the probability that the system relaxes to a lower energy state by coupling with a phonon or a photon should increase progressively. We can therefore deduce that multi-level simulations including electron-phonon coupling (or other dissipative phenomena) would be necessary to estimate $T_1$ properly. 
It is highly unlikely that electron-phonon coupling will contribute to increase $T_1$ relative to $T_{1}'$ because the leakage of the hole state to states $| \varphi_{n}^{\alpha} \rangle$ with $n>1$ is almost as effective with up $(\alpha = \uparrow)$ and down $(\alpha = \downarrow)$ spin states. So the question is whether this leakage will remain the dominant effect $(T_{1} \approx T_{1}')$ or whether the electron-phonon coupling will reduce $T_1$, which is obviously the case when phonon relaxation between $\varphi_{1}^{\uparrow}$ and $\varphi_{1}^{\downarrow}$ states becomes the dominant effect ($T_{1}^{ph}$ is typically in the $10^{-3}$--$10^{-1}$s range \cite{Li20}). More generally, this raises very interesting questions about the cross influences between charge and spin relaxations in this system.

\subsection{Comparison to experiments}

It is now important to try to compare our simulation results with experimental data. In this section, we focus on Ref.~\cite{Piot22} which reports on a four-gate device (G1-G4) fabricated from natural silicon. Remarkably, the authors of this work are able to confine a single hole under G2, which allows a more direct comparison to the theoretical simulations.

Spin coherence measurements show the existence of very low frequency noise ($10^{-4}-10^{-2}$ Hz) probably induced by both hyperfine interactions and electrical fluctuations. Further measurements were therefore performed following a Hahn echo protocol that gets rid of the low frequency noise sources. The echo amplitude follows a decay law in the form of a stretched exponential as a function of the waiting time (free evolution), representative of the high frequency noise power spectrum ($10^{4}-10^{6}$ Hz). The characteristic time which is deduced, denoted as $T_{2}^{E}$, depends on the orientation of the magnetic field and reaches the remarkable value of 88 µs at its maximum. 

These authors of Ref.~\cite{Piot22} clearly show that this high frequency noise has an electrical origin. Let us assume that it comes from a very small number of fluctuators like those studied here.  Each fluctuator $n$ can be characterized by its threshold angular frequency $\omega_{th}^{n}$ and by its oscillation frequency $\nu_{n}$. Let us also assume that each fluctuator remains in a non-Gaussian regime whatever the orientation of the magnetic field $(\nu_{n} \ll \omega_{th}^{n})$. It is then easy to show that $T_{2}^{*}$ is given by $2/\sum_{n}\nu_{n}$ (Section~III.A of the Supplemental Material \cite{supplemental}), does not depend on the magnetic field orientation and is not related to the noise spectrum $S(\omega)$ for $\omega \to 0$. We can therefore deduce that the experimental system of Ref.~\cite{Piot22} does not operate in this configuration.

The likely situation is that a significant part of the fluctuators involved in the measured noise are characterized by $\nu_{n} > \omega_{th}^{n}$, i.e. they operate in the Gaussian regime. As the measured time $T_{2}^{E}$ is relatively long, this means that the threshold angular frequencies $\omega_{th}^{n}$ are low, smaller than $\approx 10^{4}$ Hz. Since $\omega_{th} = |u_{\uparrow\uparrow}-u_{\downarrow\downarrow}|/\hbar$, we conclude that the fluctuators involved are characterized by weak coupling terms ($U$ matrix), which corresponds to defects very far from the qubit, are characterized by a weak charge displacement ($U \propto d$ where $d$ is the dipole), or have a dipole potential that is strongly screened, for example by a hole gas \cite{Piot22}. This situation seems reasonable, the existence of far fluctuators being likely given the complex and immense environment around the qubit. 

We cannot exclude that "non-Gaussian" fluctuators contribute to an isotropic noise background in Ref.~\cite{Piot22}. It could be also interesting in the future to characterize noisier devices in order to see if non-Gaussian behaviors induced by a small number of "closer" telegraphic fluctuators can be highlighted. One could also imagine very low noise situations where the influence of the distant environment is reduced but remains influenced by a few extremely slow fluctuators for which $\nu_{n} < \omega_{th}^{n}$. In these cases, the dephasing time would become totally independent of the magnetic field orientation.

\section{Conclusion}
We have simulated the spin decoherence in a hole qubit realized within a nanowire transistor in silicon-on-insulator technology. We consider the effect of a single fluctuating charge inducing telegraphic electrical noise. We show that the phase decoherence characterized by the time $T_2$ is well described in a two-level model but in a non-Gaussian regime when the fluctuator operates at a frequency lower than a threshold value $\omega_{th}$. 
The simulations show that there are operating conditions of the component, along so-called "sweet" lines, for which $\omega_{th}$ is shifted towards low frequencies which results in an increase in the minimum value of $T_2$.
However, this increase is limited due to the influence of non-diagonal coupling between the two states of opposite spin.
For the spin relaxation characterized by the time $T_1$, we show on the other hand that a multi-level model is required, due to the coupling between the ground state of the hole and many higher-energy states. 
These results highlight a rich and relatively unexpected physics for a model problem with a single fluctuator perturbing the qubit.
This shows the importance of quantum simulations including the most realistic description of the qubit.
Our study should motivate future work on the subject, in particular using a multi-level description including electron-phonon coupling as a dissipative phenomenon.

\begin{acknowledgments}
This work was supported by the French National Research Agency (ANR project MAQSi ANR-18-CE47-0007-02).
\end{acknowledgments}

\newpage

\onecolumngrid 

\section*{Supplemental Material}

\section{Details about numerical calculations}

Our numerical calculations are based on the same methodologies as in Ref.~\cite{Venitucci18}.
For convenience, we reproduce here the Appendix~D of this paper that provides details on methods and formalisms.

The potential in the device is computed with a finite volume Poisson solver assuming dielectric constants $\varepsilon_{\rm Si}=11.7$, $\varepsilon_{\rm SiO_2}=3.9$ and $\varepsilon_{\rm Si_3N_4}=7.5$. The wave functions of the qubit in this potential are then computed with a six-bands $\vec{k}\cdot\vec{p}$ model.

In bulk silicon, the six-bands $\vec{k}\cdot\vec{p}$ Hamiltonian \cite{Dresselhaus55} reads in the $\{\ket{\frac{3}{2},+\frac{3}{2}},\ket{\frac{3}{2},+\frac{1}{2}},\ket{\frac{3}{2},-\frac{1}{2}},\ket{\frac{3}{2},-\frac{3}{2}},\ket{\frac{1}{2},+\frac{1}{2}},\ket{\frac{1}{2},-\frac{1}{2}}\}$ Bloch functions basis set \cite{KP09}:
\begin{equation}
H_{\rm 6kp}=-
\begin{pmatrix}
P+Q & -S & R & 0 & \frac{1}{\sqrt{2}}S & -\sqrt{2}R \\
-S^* & P-Q & 0 & R & \sqrt{2}Q & -\sqrt{\frac{3}{2}}S & \\
R^* & 0 & P-Q & S & -\sqrt{\frac{3}{2}}S^* & -\sqrt{2}Q \\
0 & R^* & S^* & P+Q & \sqrt{2}R^* & \frac{1}{\sqrt{2}}S^* \\
\frac{1}{\sqrt{2}}S^* & \sqrt{2}Q & -\sqrt{\frac{3}{2}}S & \sqrt{2}R & P+\Delta & 0 \\
-\sqrt{2}R^* & -\sqrt{\frac{3}{2}}S^* & -\sqrt{2}Q & \frac{1}{\sqrt{2}}S & 0 & P+\Delta
\end{pmatrix}
\label{eq6kp}
\end{equation}

where:
\begin{eqnarray}
&&P=\frac{\hbar^2}{2m_0}\gamma_1\left(k_x^2+k_y^2+k_z^2\right) \\
&&Q=\frac{\hbar^2}{2m_0}\gamma_2\left(k_x^2+k_y^2-2k_z^2\right) \\
&&R=\frac{\hbar^2}{2m_0}\sqrt{3}\left[-\gamma_3\left(k_x^2-k_y^2\right)+2i\gamma_2k_xk_y\right] \label{eqR} \\
&&S=\frac{\hbar^2}{2m_0}2\sqrt{3}\gamma_3\left(k_x-ik_y\right)k_z \label{eqS} \,.
\end{eqnarray}

$k_x$, $k_y$ and $k_z$ are the components of the wave vector in the device axis set, $\gamma_1$, $\gamma_2$ and $\gamma_3$ are the Luttinger parameters, $m_0$ is the free electron mass, and $\Delta$ is the spin-orbit coupling parameter. In silicon, $\gamma_1=4.285$, $\gamma_2=0.339$, $\gamma_3=1.446$ and $\Delta=44$ meV.

The effect of the magnetic field on the Bloch functions and spin is described by the following Hamiltonian \cite{Luttinger56}:

\begin{equation}
H_{\rm Bloch}=-(3\kappa+1)\mu_B\vec{B}\cdot\vec{L}+g_0\mu_B\vec{B}\cdot\vec{S}=\mu_B\vec{B}\cdot\vec{K}\,,
\label{eqHbloch}
\end{equation}

\noindent where $\vec{L}$ is the (orbital) angular momentum of the Bloch function, $\vec{S}$ its spin, and $\kappa=-0.42$ in silicon. We neglect the effects of the much smaller $\propto q$ term of Ref.~\cite{Luttinger56}. We give below the expression of the matrices $K_x$, $K_y$, $K_z$ consistent with our choice of phases for the Bloch functions [taking $g_0=2$ in Eq.~(\ref{eqHbloch})]:
 
\begin{equation}
K_x=-
\begin{pmatrix}
0 & \sqrt{3}\kappa & 0 & 0 & -\sqrt{\frac{3}{2}}\kappa^\prime & 0 \\
\sqrt{3}\kappa  & 0 & 2\kappa & 0 & 0 & -\frac{\kappa^\prime}{\sqrt{2}} \\
0 & 2\kappa & 0 & \sqrt{3}\kappa & \frac{\kappa^\prime}{\sqrt{2}} & 0 \\
0 & 0 & \sqrt{3}\kappa & 0 & 0 & \sqrt{\frac{3}{2}}\kappa^\prime \\
-\sqrt{\frac{3}{2}}\kappa^\prime & 0 & \frac{\kappa^\prime}{\sqrt{2}} & 0 & 0 & \kappa^{\prime\prime} \\
0 & -\frac{\kappa^\prime}{\sqrt{2}} & 0 & \sqrt{\frac{3}{2}}\kappa^\prime & \kappa^{\prime\prime} & 0
\end{pmatrix} 
\end{equation}

\begin{equation}
K_y=i
\begin{pmatrix}
0 & \sqrt{3}\kappa & 0 & 0 & -\sqrt{\frac{3}{2}}\kappa^\prime & 0 \\
-\sqrt{3}\kappa  & 0 & 2\kappa & 0 & 0 & -\frac{\kappa^\prime}{\sqrt{2}} \\
0 & -2\kappa & 0 & \sqrt{3}\kappa & -\frac{\kappa^\prime}{\sqrt{2}} & 0 \\
0 & 0 & -\sqrt{3}\kappa & 0 & 0 & -\sqrt{\frac{3}{2}}\kappa^\prime \\
\sqrt{\frac{3}{2}}\kappa^\prime & 0 & \frac{\kappa^\prime}{\sqrt{2}} & 0 & 0 & \kappa^{\prime\prime} \\
0 & \frac{\kappa^\prime}{\sqrt{2}} & 0 & \sqrt{\frac{3}{2}}\kappa^\prime & -\kappa^{\prime\prime} & 0
\end{pmatrix}
\end{equation}

\begin{equation}
K_z=-
\begin{pmatrix}
3\kappa & 0 & 0 & 0 & 0 & 0 \\
0 & \kappa & 0 & 0 & \sqrt{2}\kappa^\prime & 0 \\
0 & 0 & -\kappa & 0 & 0 & \sqrt{2}\kappa^\prime \\
0 & 0 & 0 & -3\kappa & 0 & 0 \\
0 & \sqrt{2}\kappa^\prime & 0 & 0 & \kappa^{\prime\prime}  & 0 \\
0 & 0 & \sqrt{2}\kappa^\prime & 0 & 0 & -\kappa^{\prime\prime}
\end{pmatrix}\,,
\end{equation}

\noindent with $\kappa^\prime=1+\kappa$ and $\kappa^{\prime\prime}=1+2\kappa$. Note that in the $J=3/2$ subspace (the top left $4\times4$ sub-blocks of $K_x$, $K_y$ and $K_z$), $H_{\rm Bloch}$ is formally equivalent to $-2\kappa\mu_B\vec{B}\cdot\vec{J}$, where $\vec{J}=\vec{L}+\vec{S}$ is the total angular momentum of the Bloch function \cite{Luttinger56}. The eigenstates are computed with an iterative Jacobi-Davidson eigensolver \cite{Sleijpen00,Templates00}.

\section{Dependance of the perturbation matrix elements with magnetic field}

The time-dependent Hamiltonian of the system described in the main document is rewritten as
\begin{equation}
H(\vec{B},t)= H_{0}(\vec{B})+ \chi(t) U
\end{equation}
\noindent where we highlight the dependence of $H_{0}$ on the static magnetic field $\vec{B}$. The electrostatic perturbation $U(\vec{r})$ induced by the trapped charge depends neither on $\vec{B}$ nor on the electron (hole) spin.
The eigenstates of $H_{0}(\vec{B})$ are $\ket{ \varphi_{n}^{\uparrow} (\vec{B}) }$ and $\ket{ \varphi_{n}^{\downarrow} (\vec{B}) }$ with energy $E_{n}^{\uparrow} (\vec{B})$ and $E_{n}^{\downarrow} (\vec{B})$, respectively. $\uparrow$ and $\downarrow$ represent a generalized (pseudo-)spin since the physical spin is not a good quantum number in presence of spin-orbit coupling.

The decoherence processes of the spin qubit are described by the matrix of $U$ in the basis of the states $\ket{ \varphi_{1}^{\uparrow} (\vec{B}) }$ and $\ket{ \varphi_{1}^{\downarrow} (\vec{B}) }$. In this section, we discuss the evolution of this matrix with respect to $\vec{B}$, following closely the derivation of Ref.~\cite{Venitucci18}.
$H_{0}(\vec{B})$ can be expanded in powers of $\vec{B}$:
\begin{equation}
H_{0}(\vec{B}) \approx H_{0}(0) -\vec{B} \cdot \vec{M} + \mathcal{O}(B^{2})
\end{equation}
\noindent where $M_{\alpha} = -\partial H /\partial B_{\alpha}|_{B=0}$. Second and higher order terms can be safely neglected \cite{Venitucci18}.

\subsection{Case of zero magnetic field} 

The levels of the doublet are Kramers degenerate for $B=|\vec{B}|=0$, i.e., $E_{1}^{\uparrow} (0) = E_{1}^{\downarrow} (0) = E_{1}$. In addition, we can choose the phase of the wavefunctions so that
\begin{equation}
\ket{ \varphi_{1}^{\uparrow} (0) } = T \ket{ \varphi_{1}^{\downarrow} (0) }
\end{equation}
\noindent where $T$ is the time-reversal symmetry operator.

Writing $\ket{ \varphi_{1}^{\downarrow} (0) } = \alpha(\vec{r}) \ket{+} + \beta(\vec{r}) \ket{-}$ where $\ket{+}$ and $\ket{-}$ are the physical spin components, we obtain
\begin{equation}
\ket{ \varphi_{1}^{\uparrow} (0) } = T \ket{ \varphi_{1}^{\downarrow} (0) } = \beta^{*}(\vec{r}) \ket{+} - \alpha^{*}(\vec{r}) \ket{-}
\end{equation}
\noindent from which we deduce
\begin{eqnarray}
&&u_{0} = \bra{ \varphi_{1}^{\uparrow} (0) } U \ket{ \varphi_{1}^{\uparrow} (0) } = \bra{ \varphi_{1}^{\downarrow} (0) } U \ket{ \varphi_{1}^{\downarrow} (0) } = \nonumber \\
&&= \int \left[ |\alpha(\vec{r})|^{2} + |\beta(\vec{r})|^{2} \right] U(\vec{r}) d^{3}\vec{r} \label{eq_u0}
\end{eqnarray}

We used the fact that $U(\vec{r})$ does not involve the spin.
Similarly, we obtain:
\begin{equation}
\bra{ \varphi_{1}^{\uparrow} (0) } U \ket{ \varphi_{1}^{\downarrow} (0) } = \int \left[\beta(\vec{r}) \alpha(\vec{r}) - \alpha(\vec{r}) \beta(\vec{r}) \right] U(\vec{r}) d^{3}\vec{r} = 0.
\end{equation}
In absence of magnetic field, the effect of $U$ is just a rigid shift of the energy levels, and the states remain uncoupled. Time-reversal symmetry breaking is needed for a non-zero coupling \cite{Venitucci18}.

\subsection{Case of non-zero magnetic field} 

The energy splitting between the levels $n=1$ and $n=2$ being large compared to the magnetic field Hamiltonian, first-order perturbation theory can be used to derive the states for $\vec{B} \neq 0$
 \begin{eqnarray}
\ket{ \varphi_{1}^{\uparrow} (\vec{B}) } = \ket{ \varphi_{1}^{\uparrow} (0) } -B \sum_{n>1,\sigma} \frac{\bra{ \varphi_{n}^{\sigma} (0) } \vec{b} \cdot \vec{M} \ket{ \varphi_{1}^{\uparrow} (0) }}{E_{1}-E_{n}} \ket{ \varphi_{n}^{\sigma} (0) } \\ 
\ket{ \varphi_{1}^{\downarrow} (\vec{B}) } = \ket{ \varphi_{1}^{\downarrow} (0) } -B \sum_{n>1,\sigma} \frac{\bra{ \varphi_{n}^{\sigma} (0) } \vec{b} \cdot \vec{M} \ket{ \varphi_{1}^{\downarrow} (0) }}{E_{1}-E_{n}} \ket{ \varphi_{n}^{\sigma} (0) } 
 \end{eqnarray}
\noindent where $\vec{b} = \vec{B}/B$. 
Here we have chosen $\ket{ \varphi_{1}^{\uparrow} (0) }$ and $\ket{ \varphi_{1}^{\downarrow} (0) }$ so that $\bra{ \varphi_{1}^{\uparrow} (0) } \vec{b} \cdot \vec{M} \ket{ \varphi_{1}^{\downarrow} (0)} =0$ (by diagonalizing $ \vec{b} \cdot \vec{M}$ in the Kramers doublet subspace).
The non-diagonal term of the matrix $U$ can be written as
\begin{equation}
u_{\uparrow\downarrow} = \bra{ \varphi_{1}^{\uparrow} (\vec{B}) } U \ket{ \varphi_{1}^{\downarrow} (\vec{B}) } = \eta_{\uparrow \downarrow}(\vec{b}) B
\end{equation}
\noindent with
\begin{eqnarray}
\eta_{\uparrow \downarrow}(\vec{b}) = &&-\sum_{n>1,\sigma} \frac{\bra{ \varphi_{n}^{\sigma} (0) } \vec{b} \cdot \vec{M} \ket{ \varphi_{1}^{\downarrow} (0) }}{E_{1}-E_{n}} \bra{ \varphi_{1}^{\uparrow} (0) } U \ket{ \varphi_{n}^{\sigma} (0) } \nonumber \\
&&-\sum_{n>1,\sigma} \frac{\bra{ \varphi_{1}^{\uparrow} (0) } \vec{b} \cdot \vec{M} \ket{ \varphi_{n}^{\sigma} (0) }}{E_{1}-E_{n}} \bra{ \varphi_{n}^{\sigma} (0) } U \ket{ \varphi_{1}^{\downarrow} (0) }. \nonumber \\
 \end{eqnarray}
 
Similar expressions can be derived for diagonal terms:
\begin{eqnarray}
&&u_{\uparrow\uparrow} = \bra{ \varphi_{1}^{\uparrow} (\vec{B}) } U \ket{ \varphi_{1}^{\uparrow} (\vec{B}) } = u_{0} + \eta_{\uparrow \uparrow}(\vec{b}) B \nonumber \\
&&u_{\downarrow\downarrow} = \bra{ \varphi_{1}^{\downarrow} (\vec{B}) } U \ket{ \varphi_{1}^{\downarrow} (\vec{B}) } = u_{0} + \eta_{\downarrow \downarrow}(\vec{b}) B.
\end{eqnarray}

Formally similar expressions can as well be obtained for the terms $u_{n}^{\uparrow\uparrow}$ and $u_{n}^{\downarrow\downarrow}$ of the other states ($n > 1$). We deduce that the angular frequency $\omega_{th}$ that characterizes the analytic expression for $T_2$ [Eq.~(4) of the main document] is proportional to $B$,
\begin{equation}
\omega_{th} = \frac{| u_{\uparrow\uparrow}-u_{\downarrow\downarrow} |}{\hbar} = \frac{| \eta_{\uparrow\uparrow}(\vec{b})-\eta_{\downarrow\downarrow}(\vec{b}) |}{\hbar} B,
\end{equation}
\noindent and therefore the dephasing time for $\nu=\omega_{th}$ varies as $1/B$, 
\begin{equation}
T_{2}(\nu=\omega_{th}) = \frac{2\hbar}{| \eta_{\uparrow\uparrow}(\vec{b})-\eta_{\downarrow\downarrow}(\vec{b}) |B}.
\end{equation}

\section{Origin of the law in $2/\nu = 1/\nu_{cl}$ of the dephasing time $T_2$}
\subsection{General arguments}

In this section, we are interested in the dephasing time $T_2$ due to a telegraphic signal of ''classical'' frequency $\nu_{cl}=\nu/2$.
In a time interval $[0,t]$, the average number of flips is equal to $\nu_{cl}t$.
In this case, the Poisson distribution gives the probability that the fluctuator switches exactly $n$ times during the elapsed time $t$:
\begin{equation}
P_{n}(t) = \frac{(\nu_{cl}t)^{n}}{n!}\exp(-\nu_{cl}t).
\end{equation}
The probability $P_{0}(t)$ of not switching is therefore equal to $\exp(-\nu_{cl}t)$.

Consider a system characterized by the Larmor angular frequencies $\Omega$ and $\Omega'$ in states 0 and 1, respectively.
The phase shift $\delta\phi(t)$ of the qubit precession thus varies as $(\Omega'-\Omega)(t-t_{0})$ after the first switch from state 0 to state 1 at time $t_0$ ($\delta\phi(t)=0$ for $t<t_{0}$).

$T_2$ characterizes the decay of the quantity $\langle \exp(i\delta\phi(t)) \rangle_{\{E\}}$. We can write:
\begin{equation}
\langle \exp(i\delta\phi(t)) \rangle_{\{E\}} = \sum_{n} P_{n}(t) \langle \exp(i\delta\phi_{n}(t)) \rangle_{\{E_{n}\}}
\end{equation}
\noindent where $\delta\phi_{n}(t)$ represents the phase shifts in all situations $\in \{E_{n}\}$ where the fluctuator has switched exactly $n$ times during the elapsed time $t$.

We now consider the configuration where $|\Omega'-\Omega| \gg \nu$, i.e. where the dephasing angular frequency is large compared to the frequency of the telegraph noise. In this case, $\exp(i\delta\phi_{n}(t))$ averages to zero whenever there is at least one switch within $[0, t]$. Since only $\delta\phi_{0}(t)=0$, we obtain
\begin{equation}
\langle \exp(i\delta\phi(t)) \rangle_{\{E\}} \approx P_{0}(t) \langle \exp(i\delta\phi_{0}(t)) \rangle_{\{E_{0}\}} = \exp(-\nu_{cl}t)
\end{equation}
\noindent from which we deduce $T_{2} = 1/\nu_{cl} = 2/\nu$.

\modif{This calculation can be easily generalized to the situation where the qubit is influenced by $M$ fluctuators in the case where the change of angular frequency $\Omega'^{(j)}-\Omega$ induced by each fluctuator $j$ is large compared to its switching frequency $\nu^{(j)}$. The probability that the fluctuator $j$ does not switch is therefore $P_{0}^{(j)}(t) = \exp(-\nu_{cl}^{(j)}t)$ with $\nu_{cl}^{(j)} = \nu^{(j)}/2$. We obtain }

\modif{
\begin{equation}
\langle \exp(i\delta\phi(t)) \rangle_{\{E\}} \approx \prod_{j} P_{0}^{(j)}(t)  = \exp\left(-t\sum_{j}\nu_{cl}^{(j)} \right)
\end{equation}
}

\noindent \modif{from which we deduce $T_{2} =  2/\sum_{j}\nu^{(j)}$.}

\modif{In the (probable) case where one of the fluctuators is much faster than the others, $T_2$ is well given by $2/\max(\nu^{(j)})$, i.e. the coherence of the qubit is limited by the fastest of the fluctuators that perturb it. }

\subsection{Case of a two-level model}
In the two-level system discussed in the main document, the Hamiltonian in state 1 is
\begin{equation} 
H=H_{0}+U =  \left(
\begin{array}{cc}
\hbar \Omega/2 + u_{\uparrow\uparrow} &  u_{\uparrow\downarrow} \\ 
u_{\uparrow\downarrow}^{*} & -\hbar \Omega/2 + u_{\downarrow\downarrow}
\end{array} \right).
\end{equation}

After diagonalization, the Larmor angular frequency $\Omega'$ in the state 1 is therefore
\begin{equation}
\hbar \Omega' = 2\sqrt{\left( \frac{\hbar \Omega + u_{\uparrow\uparrow} - u_{\downarrow\downarrow}}{2} \right)^{2} + |u_{\uparrow\downarrow}|^2}.
\end{equation}

We define the threshold angular frequency
\begin{equation}
\omega_{th} = |\Omega' - \Omega| = \left| \frac{2}{\hbar} \sqrt{\left( \frac{\hbar \Omega + u_{\uparrow\uparrow} - u_{\downarrow\downarrow}}{2} \right)^{2} + |u_{\uparrow\downarrow}|^{2}} - \Omega \right|
\end{equation}
\noindent which, in the pure dephasing model $|u_{\uparrow\uparrow} - u_{\downarrow\downarrow}| \gg |u_{\uparrow\downarrow}|$, gives Eq.~(3) of the main document, i.e. $\omega_{th} \approx |u_{\uparrow\uparrow} - u_{\downarrow\downarrow}|/\hbar$ (because $|u_{\uparrow\uparrow} - u_{\downarrow\downarrow}| \ll \hbar \Omega$).

In the opposite case where $|u_{\uparrow\uparrow} - u_{\downarrow\downarrow}| \ll |u_{\uparrow\downarrow}|$, the threshold angular frequency becomes
\begin{equation}
\omega_{th} \approx \frac{2|u_{\uparrow\downarrow}|^{2}}{\hbar^{2}\Omega}
\end{equation}
\noindent which is valid in particular when one seeks to reach a "sweet" point where $u_{\uparrow\uparrow} - u_{\downarrow\downarrow} \to 0$. Remarquably, $\omega_{th}$ can be rewritten in this case as $4/T_{1}^{\rm min}$ where $T_{1}^{\rm min}$ is the minimum value of $T_1$ in the two-level model [Eq.~(5) of the main document].

In this section, we thus conlude that, for low-frequency telegraphic noise such that $\nu \ll \omega_{th}$, the dephasing time is always given by the universal law $T_{2} = 2/\nu$.

\section{$g$ matrix of the hole qubit}
\subsection{General model}

The first-order expansion of the Hamiltonian in $\vec{B}$ described in the previous section can be rewritten in a general way in the $g$-matrix formalism as
\begin{equation}
H_{0}(\vec{B},V_{BG}) = H_{0}(0,V_{BG}) + H_{Z}(\vec{B},V_{BG})
\end{equation}
\noindent with a Zeeman Hamiltonian given by
\begin{equation}
H_{Z}(\vec{B},V_{BG}) = \frac{1}{2} \mu_{B} \boldsymbol{\sigma} \cdot \hat{g}(V_{BG}) \cdot \vec{B}
\end{equation}
\noindent where $\boldsymbol{\sigma}$ is the vector of Pauli matrices, and $\hat{g}(V_{BG})$ is a real $3 \times 3$ matrix. Here we highlight the dependence of the Hamiltonian and the $g$-matrix on the back-gate bias $V_{BG}$. The dependence on another potential could be considered in the same way. We have assumed that the Hamiltonian is written in some basis $\{\ket{\Uparrow}, \ket{\Downarrow}\}$ in which the vectors are orthogonal linear combinations of $\ket{ \varphi_{1}^{\uparrow} (0) }$ and $\ket{ \varphi_{1}^{\downarrow} (0) }$. The $\hat{g}$ matrix is not unique, as it depends on the choice of the magnetic field axes and of the hole states basis. 

As $xy$ is an exact mirror symmetry plane of the device whatever $V_{BG}$, and $yz$ is an approximate mirror symmetry plane, $x$, $y$ and $z$ can be considered as the principal magnetic axes of the system \cite{Venitucci18}.
For a magnetic field in the $xy$ plane, $\vec{B} = B\,(\cos(\varphi),\sin(\varphi),0)$, the Zeeman Hamiltonian can hence be written as
\begin{equation}
H_{Z}(\vec{B},V_{BG})  =  \frac{1}{2} \mu_{B} B \left[ \begin{array}{cc}
g_{x} \cos(\varphi)   & -i g_{y} \sin(\varphi) \\
+i g_{y} \sin(\varphi)  & -g_{x} \cos(\varphi)
\end{array} \right] \label{eqn_matrix_g}
\end{equation}
\noindent in which $g_x$ and $g_y$ depend implicitely on $V_{BG}$. The Zeeman splitting is equal to
\begin{equation}
\hbar \Omega = \mu_{B} B \sqrt{g_{x}^{2} \cos^{2}(\varphi) + g_{y}^{2} \sin^{2}(\varphi)}.
\label{eqn_Zeeman_splitting}
\end{equation}

\begin{figure}%[!h]
\centering
\includegraphics[width=0.6\columnwidth]{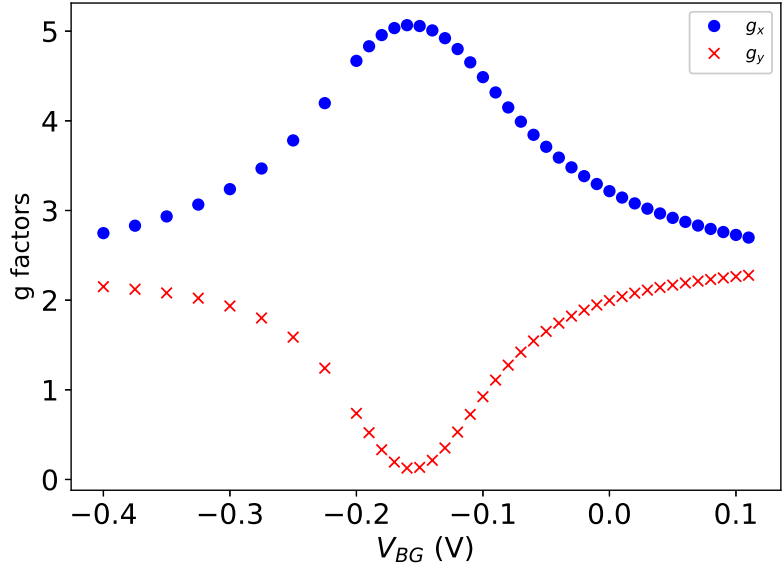}
\caption{$g_x$ and $g_y$ factors for the device considered in the present work.}
\label{fig_cg_factors}
\end{figure}

The $g_x$ and $g_y$ factors calculated for the device studied in this work are shown in Fig.~\ref{fig_cg_factors}. 
Very similar $g$ factors were obtained for a nearly identical device and were discussed in Ref.~\cite{Venitucci18}.
The evolution of these factors with $V_{BG}$ reflects the respective weight of the wave function on the heavy and light hole Bloch functions. For $V_{BG}\approx -0.15$~V, the wave function is located at the center of the nanowire, the heavy hole component is maximal as in a (100) silicon film. 
In this case, the in-plane $g$ factors are almost zero, and the vertical $g$ factor $(g_{x})$ is maximum.
For higher or smaller values of $V_{BG}$, the wave function is pushed to one side of the nanowire. The strong lateral confinement induces a significant increase in the weight of the wave function on the light hole states.
As shown in Ref.~\cite{Piot22}, this transfer of weight between heavy and light hole states makes that (Fig.~\ref{fig_cg_factors})
\begin{equation}
\frac{\partial g_{x}}{\partial V_{BG}} \approx - \frac{\partial g_{y}}{\partial V_{BG}}.
\label{eq_partial_vbg}
\end{equation}

It is interesting to find the conditions for which the Zeeman splitting [Eq.~(\ref{eqn_Zeeman_splitting})] is stationary with respect to $V_{BG}$.
Using Eq.~(\ref{eq_partial_vbg}), we deduce that $(\partial \,\hbar \Omega) / (\partial \,V_{BG}) = 0$ for
\begin{eqnarray}
&&g_{x} \cos^{2}\varphi = g_{y} \sin^{2}\varphi \label{eqn_vertical_lines_1} \\
&&\varphi \approx \frac{\pi}{2} \pm \arctan \sqrt{\frac{g_x}{g_y}}. \label{eqn_vertical_lines_2}
\end{eqnarray}

It is also important to remind that the dependence of the Zeeman Hamiltonian and $g$ factors on $V_{BG}$ is essentially through the electric field along the $y$ axis. Similar results can be obtained by playing on the potential of the other gates, only the lever arm will be different \cite{Piot22}.

\subsection{Diagonal matrix element of the perturbation}
The eigenstates of the Hamiltonian given by Eq.~(\ref{eqn_matrix_g}) are 
\begin{eqnarray}
&&\ket{\varphi_{1}^{\uparrow}} = \alpha \ket{\Uparrow} + \beta \ket{\Downarrow} \\
&&\ket{\varphi_{1}^{\downarrow}} = -\beta \ket{\Uparrow} + \alpha^{*} \ket{\Downarrow}
\end{eqnarray}
\noindent with
\begin{eqnarray}
&&\alpha =  \frac{i b_{y} g_{y}}{\sqrt{b_{y}^{2} g_{y}^{2} + \left(b_{x} g_{x} + \sqrt{b_{x}^{2} g_{x}^{2} + b_{y}^{2} g_{y}^{2}}\right)^{2}}} \\
&&\beta =  \frac{b_{x} g_{x} + \sqrt{b_{x}^{2} g_{x}^{2} + b_{y}^{2} g_{y}^{2}}}{\sqrt{b_{y}^{2} g_{y}^{2} + \left(b_{x} g_{x} + \sqrt{b_{x}^{2} g_{x}^{2} + b_{y}^{2} g_{y}^{2}}\right)^{2}}}
\end{eqnarray}
\noindent in which we may write $b_{x} = \cos(\varphi)$ and $b_{y} = \sin(\varphi)$, for simplicity.

The effect of a fluctuating electric charge results in a perturbation Hamiltonian 
\begin{equation}
U = \delta H_{0}(0,V_{BG}) + \delta H_{Z}(\vec{B},V_{BG})
\end{equation}
\noindent with
\begin{equation}
\delta H_{Z}  =  \frac{1}{2} \mu_{B} B \left[ \begin{array}{cc}
\delta g_{x} b_{x}   & -i \delta g_{y} b_{y} \\
+i \delta g_{y} b_{y}  & -\delta g_{x} b_{x}
\end{array} \right]  + \delta H'_{Z} \label{eqn_matrix_dg}
\end{equation}
\noindent in which, for the reason discussed above, we have $\delta g_{x} \approx -\delta g_{y}$. 
$\delta H'_{Z}$ contains additional terms like $\delta g_{xy} b_{y}, \delta g_{yx} b_{x}$ arising from the fact that the variation (differential) of the $g$ matrix may not be diagonal in the same basis and magnetic axes frame as $\hat{g}$ \cite{Venitucci18}.
However, in the present case, these terms are small and the dephasing process is mainly determined by the diagonal matrix elements $u_{\uparrow\uparrow} \approx \bra{\varphi_{1}^{\uparrow}} \delta H_{0} + \delta H_{Z} \ket{\varphi_{1}^{\uparrow}}$ and $u_{\downarrow\downarrow} \approx \bra{\varphi_{1}^{\downarrow}} \delta H_{0} + \delta H_{Z} \ket{\varphi_{1}^{\downarrow}}$.
After some algebra, we obtain:
\begin{eqnarray}
u_{\uparrow\uparrow} \approx u_{0} + \delta u \\
u_{\downarrow\downarrow} \approx u_{0} - \delta u
\end{eqnarray}
\noindent with $u_{0} = \bra{\varphi_{1}^{\uparrow}} \delta H_{0} \ket{\varphi_{1}^{\uparrow}} = \bra{\varphi_{1}^{\downarrow}} \delta H_{0} \ket{\varphi_{1}^{\downarrow}} $ [see Eq.~(\ref{eq_u0})] and
\begin{widetext}
\begin{equation}
\delta u = - \frac{\delta g_{x} b_{x}^{3} g_{x}^{2} + \delta g_{x} b_{x}^{2} g_{x} \sqrt{b_{x}^{2} g_{x}^{2} + b_{y}^{2} g_{y}^{2}} + \delta g_{y} b_{x} b_{y}^{2} g_{x} g_{y} + \delta g_{y} b_{y}^{2} g_{y} \sqrt{b_{x}^{2} g_{x}^{2} + b_{y}^{2} g_{y}^{2}}}{b_{x}^{2} g_{x}^{2} + b_{x} g_{x} \sqrt{b_{x}^{2} g_{x}^{2} + b_{y}^{2} g_{y}^{2}} + b_{y}^{2} g_{y}^{2}}.
\end{equation}
\end{widetext}

We can verify that $\delta u$ and therefore $u_{\uparrow\uparrow} - u_{\downarrow\downarrow}$ cancels for $\delta g_{x} = -\delta g_{y}$ and $g_{x}b_{x}^{2} = g_{y}b_{y}^{2}$ [Eq.~(\ref{eqn_vertical_lines_1})], i.e. when $(\partial \,\hbar \Omega) / (\partial \,V_{BG}) = 0$. 

\section{Long-time limit of a $N$-level system perturbed by a low-frequency telegraphic noise}
\subsection{General model}

We consider a system of $N$ levels $\{ E_{1}=\hbar \omega_{1} \cdots E_{N}=\hbar \omega_{N} \}$ and wavefunctions $\{\ket{\phi_{1}} \cdots \ket{\phi_{N}} \}$, eigenvalues and eigenstates of a Hamiltonian $H_0$. The system is influenced by a telegraphic noise that fluctuates between two configurations ``0'' and ``1''. The total Hamiltonian is therefore $H(t) = H_{0}+\chi(t) U$, where $U$ is the perturbation when the system switches from 0 to 1 and $\chi(t)=0,1$. The eigenvalues of $H_{0}+U$ are labelled $E'_{i}=\hbar \omega'_{i}$.
The system switches between the two configurations at the times $t_{1}, t_{2} \cdots$.

We must solve the time-dependent Schrödinger equation:
\begin{equation}
i\hbar \frac{d \Psi(t)}{d t} = H(t) \Psi(t) 
\label{H_t}
\end{equation}

We assume that $H(t) = H_{0}$ for $t_{0} \leq t < t_{1}$ with the initial condition $\Psi(t_{0}) = \Psi_{0}$. The propagation of the state can be easily written. For example, since the system is in the configuration 0 between $t_0$ and $t_1$ and the configuration 1 between $t_1$ and $t_2$, the wavefunction at $t_{2}$ is given by:
\begin{equation}
P T'(\Delta t_{2}) P^{+} T(\Delta t_{1})  \ket{\Psi_{0}}
\label{eq_propagation}
\end{equation}
\noindent in which $T(\Delta t_{n})$ is the diagonal matrix
\begin{equation}
\left[ \begin{array}{ccc}
\exp \left(- i\omega_{1} \Delta t_{n} \right)   &  & \mbox{\huge 0}  \\
 & \ddots &  \\
\mbox{\huge 0} &  & \exp \left(- i\omega_{N} \Delta t_{n} \right)
\end{array} \right]
\end{equation}
\noindent with $\Delta t_{n} = t_{n}-t_{n-1}$. $T'(\Delta t_{n})$ is the same matrix in which the angular frequencies $\omega_{i}$ are replaced by $\omega'_{i}$. $P$ is the basis change matrix, that is, the matrix of the eigenvectors of $H_{0}+U$ in the basis of the eigenstates of $H_0$. In Eq.~(\ref{eq_propagation}), we used $P^{-1} = P^{+}$.

We can therefore proceed by recursion and define a sequence of states 
\begin{equation}
\begin{array}{c}
\ket{\Psi_{n}} = P^{+} T(\Delta t_{n}) \ket{\Psi_{n-1}} \, \textrm{for}\; n \,\textrm{odd}  \\
\ket{\Psi_{n}} = P T'(\Delta t_{n}) \ket{\Psi_{n-1}} \, \textrm{for}\; n \,\textrm{even}.
\end{array}
\label{eq_psin_psinm1}
\end{equation}

We introduce the density operator $\rho_{n} = \overline{\ket{\Psi_{n}} \bra{\Psi_{n}}}$ in which the overline means the statistical average on the different random realizations of the time interval $\Delta t_{n}$ for the mean switching frequency $\nu$.

Here we want to understand what happens in the long run, after a large number of switches of the fluctuator, in the case of a low-frequency telegraphic noise. We consider the situation where
\begin{equation}
\nu \ll \omega_{i}, \forall i.
\end{equation}

In this case, the quantities $\exp \left( i\omega_{i} t \right)$ present in the propagators can be written as $\exp \left( i\theta \right)$ where $\theta$ can be seen as a random variable between 0 and $2\pi$.

Using Eq.~(\ref{eq_psin_psinm1}), the diagonal term $(\rho_{n})_{ii}$ which gives the electronic population on the level $i$ after $n$ switches is given, for odd $n$, by
\begin{equation}
\overline{\sum_{kl} P^{+}_{ik} \exp \left(- i\omega_{k} \Delta t_{n} \right)  (\rho_{n-1})_{kl} \exp \left(+ i\omega_{l} \Delta t_{n} \right) P_{li}} .
\end{equation}

For $k \neq l$, assuming $\omega_{k} \neq \omega_{l}$ which is likely in presence of a magnetic field that splits the spin doublets, we have
\begin{equation}
\overline{ \exp \left(i(\omega_{l}-\omega_{k}) \Delta t_{n} \right)  } = 0.
\end{equation}

We deduce that
\begin{equation}
(\rho_{n})_{ii} = \sum_{k} |P_{ik}|^{2} (\rho_{n-1})_{kk},
\end{equation}
\noindent which can be rewritten as  
\begin{equation}
(\rho_{n}) = A (\rho_{n-1})
\label{eq_recursion_rho}
\end{equation}
\noindent in which $()$ represents the column vector formed by the diagonal matrix elements of the density operator and 
\begin{equation}
A = \left[ \begin{array}{ccc}
|P_{11}|^{2}  & \cdots & |P_{1N}|^{2}  \\
\vdots & \ddots & \vdots \\
|P_{N1}|^{2}  & \cdots & |P_{NN}|^{2}
\end{array} \right]. \label{eqn_matrix_A}
\end{equation}

A similar result is obtained for even values of $n$.

Let $X$ be an eigenvector of the matrix $A$ for the eigenvalue $\lambda$, and $x_i$ its largest component (in modulus).
We have:
\begin{eqnarray}
&&\sum_{j} |P_{ij}|^{2} x_{j} = \lambda x_{i} \label{eq_Pij_xj} \\
&&\implies \lambda = \sum_{j} |P_{ij}|^{2} (x_{j}/x_{i}) \label{eq_Pij_xjxi} \\ 
&&\implies  |\lambda| \leq \sum_{j} |P_{ij}|^{2} |x_{j}/x_{i}| \leq \sum_{j} |P_{ij}|^{2} = 1. \label{eq_Pij_1}
\end{eqnarray}

This shows that the eigenvalues $\lambda_{i}$ of $A$ have a modulus smaller or equal to 1.
In addition, 1 is always a trivial eigenvalue for the eigenvector
\begin{equation}
\left( \begin{array}{c} 1/\sqrt{N} \\ \vdots \\ 1/\sqrt{N} \end{array} \right).
\label{eq_vect_final}
\end{equation}

Using Eq.~(\ref{eq_Pij_xjxi}) and Eq.~(\ref{eq_Pij_1}), we further deduce that an eigenvector $X$ of $A$ for $\lambda=1$ must have components of the form $x_{j}=\exp(i\psi_{j})/\sqrt{N}$.
Injecting this into Eq.~(\ref{eq_Pij_xjxi}) and taking the complex conjugate, we obtain
\begin{eqnarray}
|\lambda|^{2} &&= \sum_{j,k} |P_{ij}|^{2} |P_{ik}|^{2} \exp \left[i(\psi_{j}-\psi_{k}) \right] \\
&&= \sum_{j,k} |P_{ij}|^{2} |P_{ik}|^{2} \cos(\psi_{j}-\psi_{k}) = 1. \label{eq_Pij_Pik}
\end{eqnarray}

We deduce that $\psi_{j}=\psi_{k}$ must be imposed for all $j$ and $k$ to verify the last equation (for $P_{ij}$ and $P_{ik}$ nonzero, see below).
Equation~(\ref{eq_vect_final}) is therefore the only eigenvector for the eigenvalue $\lambda=1$. In reality, there is an exception to this rule when the Hamiltonian matrix $H_{0}+U$ can be split into independent blocks so that $A$ is also a block diagonal matrix, i.e., with non diagonal blocks where $P_{ij}=0$. In this case, the eigenvalue $\lambda=1$ exists for each of the blocks, and $N$ must be replaced by the size of the block. 

From Eq.~(\ref{eq_recursion_rho}), we deduce the long time limit of the density:
\begin{equation}
 (\rho_{\infty}) = \lim_{n \to \infty} (\rho_{n}) = \left[ \lim_{n \to \infty} A^{n} \right] (\rho_{0})
\label{eq_lim_rho}
\end{equation}

In the basis of the eigenvectors of $A$, putting the vector given by Eq.~(\ref{eq_vect_final}) first, we have 
\begin{equation}
A_ {\infty} = \left[ \lim_{n \to \infty} A^{n} \right] = \left[ \begin{array}{cccc}
1 & 0 & \cdots & 0  \\
0 & 0 & \cdots & 0 \\
\vdots & \vdots & \ddots & \vdots \\
0 & 0 & \cdots & 0
\end{array} \right]
\label{eq_lim_An}
\end{equation}
\noindent from which we deduce using Eq.~(\ref{eq_lim_rho})
\begin{equation}
 (\rho_{\infty}) = Q A_ {\infty} Q^{-1} (\rho_{0})
\end{equation}
\noindent where $Q$ is the matrix of the eigenvectors of $A$ in which $Q_{i1}=1/\sqrt{N}$ for all $i$ [Eq.~(\ref{eq_vect_final})]. 

Similarly, we can show that the first row of the matrix $Q^{-1}$ is such that $Q^{-1}_{1i}=1/\sqrt{N}$ for all $i$ because of the orthogonality of the first column vector of $Q$ to all other column vectors. This can be deduced from Eq.~(\ref{eq_Pij_xj}) which gives
\begin{equation}
\sum_{i,j} |P_{ij}|^{2} x_{j} = \sum_{j} x_{j} = \lambda \sum_{i} x_{i}
\end{equation}
\noindent which implies that $\sum_{i} x_{i}=0$ for $\lambda \neq 1$.
We conclude that not only the eigenvalue $\lambda=1$ is nondegenerate but also that the eigenvectors associated with the other eigenvalues form an orthogonal subspace.

Using these results and Eq.~(\ref{eq_lim_An}), we deduce finally:
\begin{equation}
 (\rho_{\infty})_{i} = \sum_{j} Q_{i1} Q^{-1}_{1j} (\rho_{0})_{j} = \frac{1}{N} \sum_{j} (\rho_{0})_{j} = \frac{1}{N}.
 \label{eq_rho_final}
\end{equation}

This shows that the system always ends up in a situation of equipartition between all states of the basis, whatever the starting point.
In this proof, we have made the assumption that the matrix $A$ is diagonalizable if an eigenvalue is degenerate. If this is not the case, the result can be generalized to the case of Jordan normal forms.

\subsection{Application to the calculation of the relaxation lifetime $T_1$ in the two-level model}
As in the main document, the Hamiltonian is written as
\begin{equation} 
H(t)=  \left(
\begin{array}{cc}
-\Delta E &  0 \\ 
0 & \Delta E
\end{array} \right)
+ \chi(t) \left(
\begin{array}{cc}
0 &  u_{\uparrow\downarrow} \\ 
u_{\uparrow\downarrow}^{*} & 0 
\end{array} \right)
\end{equation}
\noindent in which we define $\Delta E = \hbar \Omega/2$. Here we assume $u_{\uparrow\uparrow}=u_{\downarrow\downarrow}=0$, and $|u_{\uparrow\downarrow}|/  \Delta E \ll 1$.

The matrix $A$ defined in Eq.~(\ref{eqn_matrix_A}) is given by
\begin{equation}
A = \left[ \begin{array}{cc}
\frac{1}{2} \left( 1 + \frac{\Delta E}{\sqrt{\Delta E^{2} + |u_{\uparrow\downarrow}|^{2}}} \right) & \frac{1}{2} \left( 1 - \frac{\Delta E}{\sqrt{\Delta E^{2} + |u_{\uparrow\downarrow}|^{2}}} \right)  \\
\frac{1}{2} \left( 1 - \frac{\Delta E}{\sqrt{\Delta E^{2} + |u_{\uparrow\downarrow}|^{2}}} \right)  & \frac{1}{2} \left( 1 + \frac{\Delta E}{\sqrt{\Delta E^{2} + |u_{\uparrow\downarrow}|^{2}}} \right)
\end{array} \right]. \label{eqn_matrix_A_22}
\end{equation}

We have:
\begin{equation}
\left( \begin{array}{c}
\rho_{\uparrow}^{n}  \\
\rho_{\downarrow}^{n}
\end{array} \right)
= A
\left( \begin{array}{c}
\rho_{\uparrow}^{n-1}  \\
\rho_{\downarrow}^{n-1}
\end{array} \right) \label{eqn_rn_A_rnm1}
\end{equation}
\noindent in which $\rho_{\uparrow}^{n}$ is the population on the state $\uparrow$ after $n$ switches.

The relaxation of the spin is described by the decay of $\sigma_{z}^{n}=\rho_{\uparrow}^{n}  - \rho_{\downarrow}^{n}$. Using Eq.~(\ref{eqn_rn_A_rnm1}), we deduce
\begin{equation}
\sigma_{z}^{n} = \frac{\Delta E}{\sqrt{\Delta E^{2} + |u_{\uparrow\downarrow}|^{2}}} \sigma_{z}^{n-1}.
\end{equation}

Using the initial condition $\sigma_{z}^{0} = 1$, we obtain
\begin{equation}
\sigma_{z}^{n} = \left(  \frac{\Delta E}{\sqrt{\Delta E^{2} + |u_{\uparrow\downarrow}|^{2}}} \right)^{n}.
\end{equation}

The average elapsed time for $n$ steps being $t_{n} = n/\nu_{cl} = 2n/\nu$, $\sigma_{z}^{n}$ can be rewritten as $\exp(-t_{n}/T_{1})$ with
\begin{equation}
T_{1} = \frac{4 \Delta E^{2}}{\nu |u_{\uparrow\downarrow}|^{2}} = \frac{\hbar^{2} \Omega^{2}}{\nu |u_{\uparrow\downarrow}|^{2}}.
\end{equation}

This result, obtained using $|u_{\uparrow\downarrow}|/  \Delta E \ll 1$, coincides with Eq.~(7) of the main document in the limit $\nu \ll \Omega$. In this case, $T_1$ was derived from the noise spectral density at frequency $\Omega$ \cite{Paladino14}.

\section{Energy levels and coupling strengths for the Traps 2 and 3}

\begin{figure}%[!h]
\centering
\includegraphics[width=0.70\columnwidth]{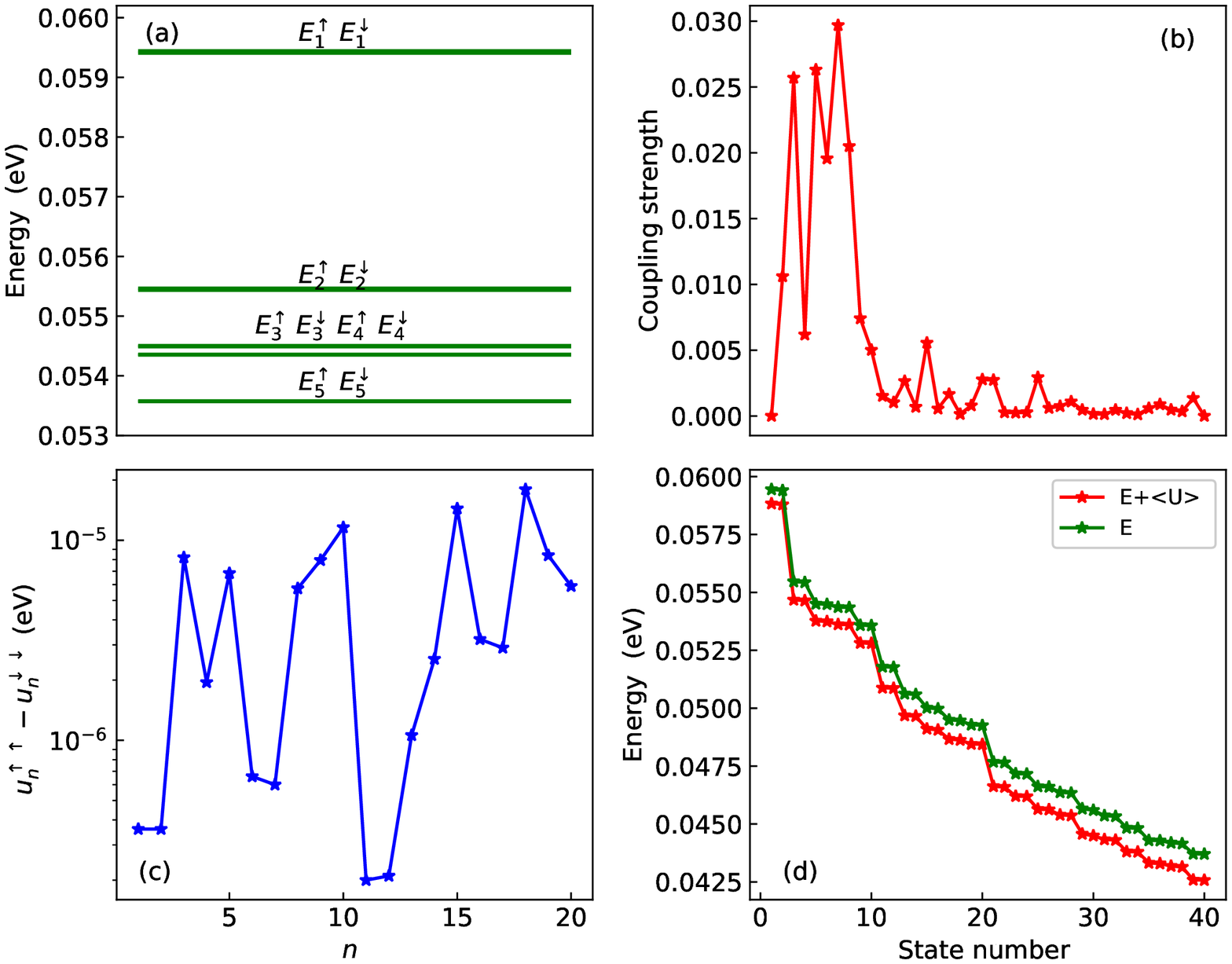}
\caption{(a) Highest electronic energy levels calculated for the hole qubit. 
(b) Coupling strength defined as the ratio $\left|\langle \varphi_{1}^{\uparrow} | U | \varphi_{n}^{\uparrow\downarrow} \rangle \right|/|E_{1}^{\uparrow} - E_{n}^{\uparrow\downarrow}|$. 
(c) $\delta_{n} = \langle \varphi_{n}^{\uparrow} | U | \varphi_{n}^{\uparrow} \rangle - \langle \varphi_{n}^{\downarrow} | U | \varphi_{n}^{\downarrow} \rangle =  u_{n}^{\uparrow\uparrow} - u_{n}^{\downarrow\downarrow}$ versus $n$.
(d) Unperturbed level energies $E_{n}^{\uparrow\downarrow}$ (green) and perturbed level energies $E_{n}^{\uparrow\downarrow}+\langle \varphi_{n}^{\uparrow\downarrow} | U | \varphi_{n}^{\uparrow\downarrow} \rangle$ (red) presented according to the state number defined as $2n-1$ for $|\varphi_{n}^{\uparrow}\rangle$ states and $2n$ for $|\varphi_{n}^{\downarrow}\rangle$ states. (b-d) All results are for Trap 2.}
\label{fig_coupling_strengths_2}
\end{figure}

\begin{figure}%[!h]
\centering
\includegraphics[width=0.70\columnwidth]{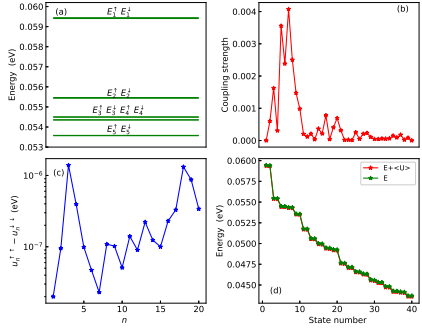}
\caption{Same as Fig.~\ref{fig_coupling_strengths_2} for Trap 3.}
\label{fig_coupling_strengths_3}
\end{figure}

Figures~\ref{fig_coupling_strengths_2} and \ref{fig_coupling_strengths_3} present information on the energy levels and on the elements of the perturbation matrix $U$ for Trap 2 and Trap 3.

\section{2D maps of $\Omega$ and $u_{\uparrow\uparrow}-u_{\downarrow\downarrow}$ in the $(\theta,\varphi)$ plane}

\begin{figure*}%[!h]
\centering
\includegraphics[width=1.5\columnwidth]{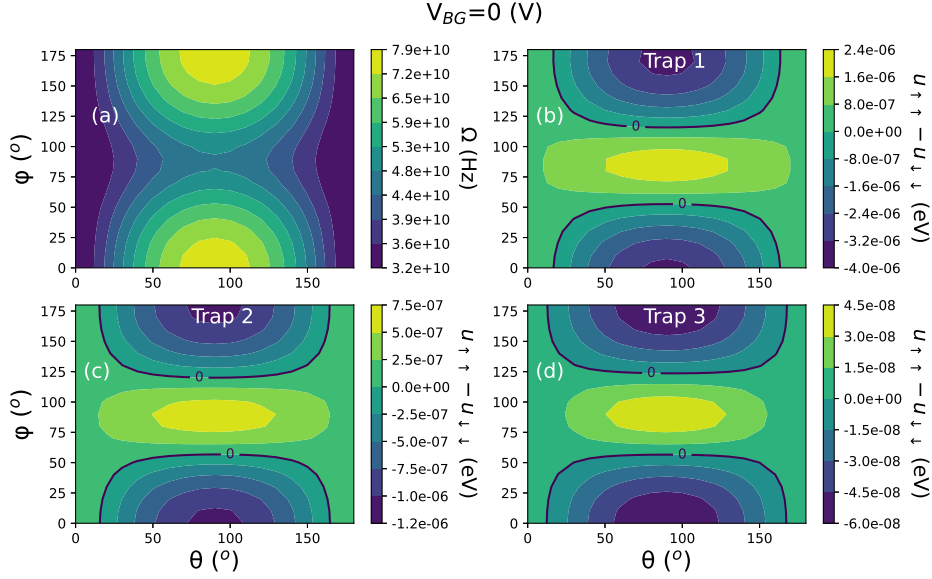}
\caption{2D plots of the Larmor frequency $\Omega$ (a) and $u_{\uparrow\uparrow}-u_{\downarrow\downarrow}$ (b-d) versus $\theta$ and $\varphi$, for Trap 1 (b), Trap 2 (c) and Trap 3 (d), for $V_{BG}=0$~V. The contours corresponding to $u_{\uparrow\uparrow}-u_{\downarrow\downarrow}=0$ are indicated by black lines. }
\label{fig_2D_map_Vbg0}
\end{figure*}

Figure~\ref{fig_2D_map_Vbg0} presents a 2D plot of $\Omega$ and $u_{\uparrow\uparrow}-u_{\downarrow\downarrow}$ versus the orientation angles of the magnetic field. 
The general behaviors are close for the three traps. In particular, the sweet lines $(u_{\uparrow\uparrow}-u_{\downarrow\downarrow} = 0)$ at $\varphi \approx 90 \pm 35^{\circ}$ remain well marked.
This demonstrates that the analysis made in the main document for a centered fluctuator close to the qubit (Trap 1) remains approximately valid for off-center and more distant fluctuators.

%
%In the case of Trap 1, the figure of $u_{\uparrow\uparrow}-u_{\downarrow\downarrow}$ is symmetric with respect to $\theta=90^{\circ}$, because $xy$ is a mirror plane of symmetry of the system. This symmetry is partially broken in the case of Traps 2 and 3, because of their position ($z$ not zero).

\section{Additional plots of $m(t)$ at different frequencies $\nu$}
\begin{figure*}%[!h]
\centering
\includegraphics[width=1.5\columnwidth]{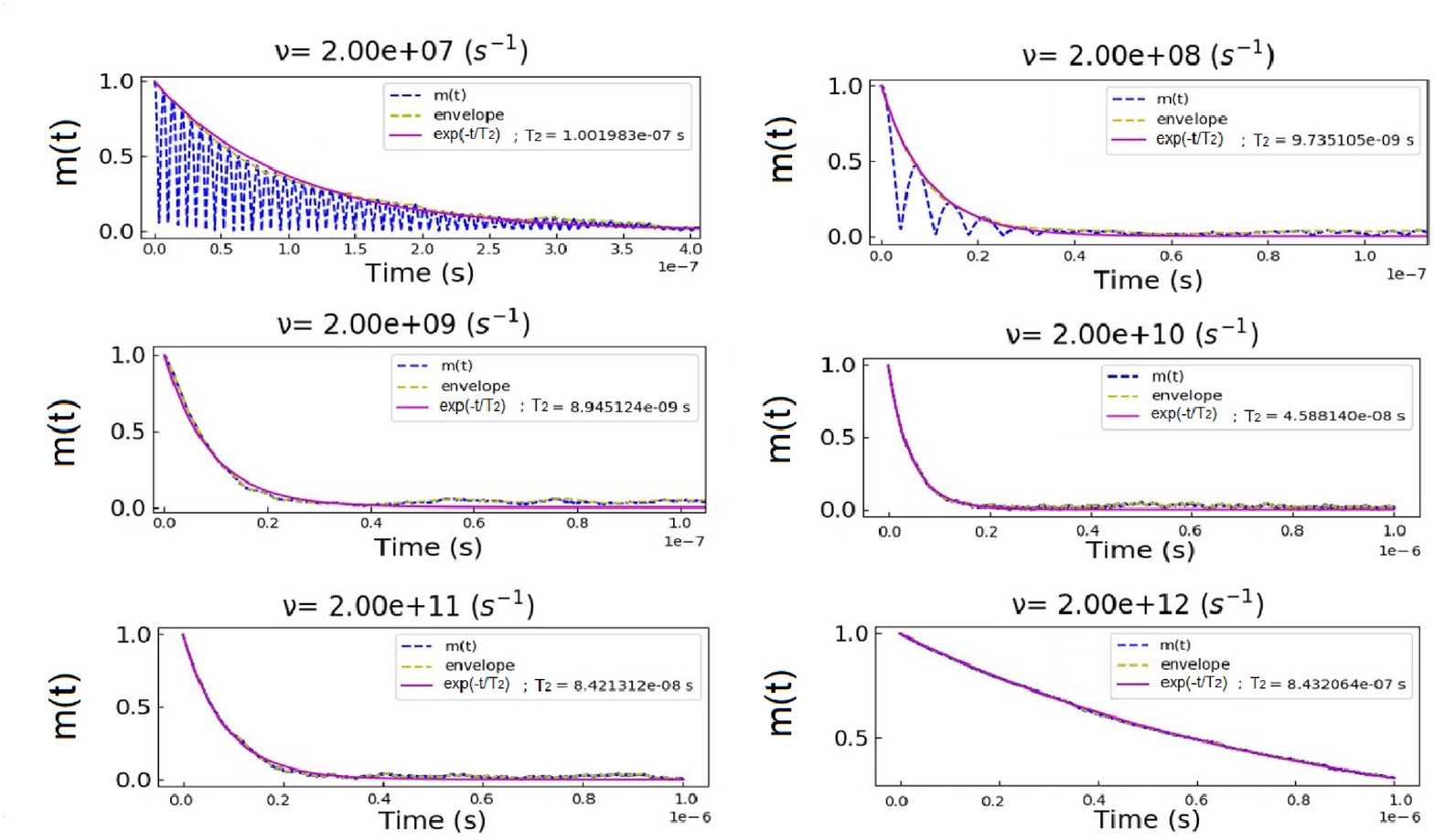}
\caption{Evolution of $m(t)$ for $\nu=2 \times 10^{7}$~s$^{-1}$, $2 \times 10^{8}$~s$^{-1}$, $2 \times 10^{9}$~s$^{-1}$, $2 \times 10^{10}$~s$^{-1}$, $2 \times 10^{11}$~s$^{-1}$ and $2 \times 10^{12}$~s$^{-1}$ (Trap~1). For $\nu < \omega_{th} = 1.963 \times 10^{9}$~s$^{-1}$, $m(t)$ presents oscillations \cite{Paladino14}. In this case, $T_2$ is given by the exponential decay of the envelope.}
\label{fig_m_t}
\end{figure*}

Figure~\ref{fig_m_t} presents additional plots of $m(t)$ for different values of $\nu$, for Trap 1. 

\appendix

\section{Calculation of the decoherence time $T_{1}'$}
\label{appendix_cal_t1}
The characteristic time $T_1'$ is defined to describe the leakage of the wave function of the hole out of the doublet composed of $\varphi_{1}^{\uparrow}$ and $\varphi_{1}^{\downarrow}$ states under the effect of the random variations of the perturbing potential. 
Our goal is to deduce $T_1'$ from the evolution of $\psi(t)$ obtained by solving the time dependent Schrödinger equation but we have to take into account two intrinsic limitations of this approach. First, a classical noise is considered while quantum effects can become important when $kT$ is small compared with the energy gaps between levels. Second, dissipation effects are not included, for example by coupling with phonons or by feedback to the fluctuator and the electron reservoirs. 
As a consequence, the long time limit of $\psi(t)$ cannot be physically correct in our model. Indeed, when the non-diagonal elements of the $U$ matrix are zero, we show in Sect.~V of the Supplemental Material \cite{supplemental} that the simulations will always converge to the situation where $\psi(t)$ is statistically uniformly distributed over all the states of the system, that is to say that $\left\langle |\langle \varphi_{n}^{\uparrow\downarrow} |\psi(t) \rangle|^2 \right\rangle_{\{E\}}$ converges to $1/(2N)$ whatever $N$ is, whereas it should tend to the value given by a quasi Fermi-Dirac statistic if dissipative phenomena were taken into account.

As a consequence, we have limited our analysis to short times for which the quantity $p_{1}(t) = \left\langle |\langle \varphi_{1}^{\uparrow} |\psi(t) \rangle|^2 + |\langle \varphi_{1}^{\downarrow} |\psi(t) \rangle|^2 \right\rangle_{\{E\}}$ has an exponential decay of the form $\exp (-t/T_{1}')$ where $T_{1}'$ quickly becomes independent of the number of states considered in the basis.
This approach is sufficient to describe the initial evolution of the wave function of the hole to the $\varphi_{n}^{\uparrow}$ and $\varphi_{n}^{\downarrow}$ states for $n>1$ and not its subsequent evolution.
 
In practice, we found the mono-exponential character of $p_{1}(t)$ when we consider the time span for which this quantity varies from 1 to 0.5, and that the value of $T_{1}'$ is converged for $N=20$ (Fig.~\ref{fig_fit_lifetimes}a). This value can easily be understood since the matrix elements of $U$ between the states $\varphi_{1}^{\uparrow\downarrow}$ and $\varphi_{n}^{\uparrow\downarrow}$ decrease sharply for increasing values of $n>10$ (Fig.~\ref{fig_levels_couplings}).

\section{Effect of non-instantaneous transitions}
\label{sect_diabatic}

The telegraphic noise model assumes that the transitions between the two states of the fluctuator are instantaneous. Here we discuss the influence of non-instantaneous transitions and their realism.

\begin{figure}%[!h]
\centering
\includegraphics[width=0.99\columnwidth]{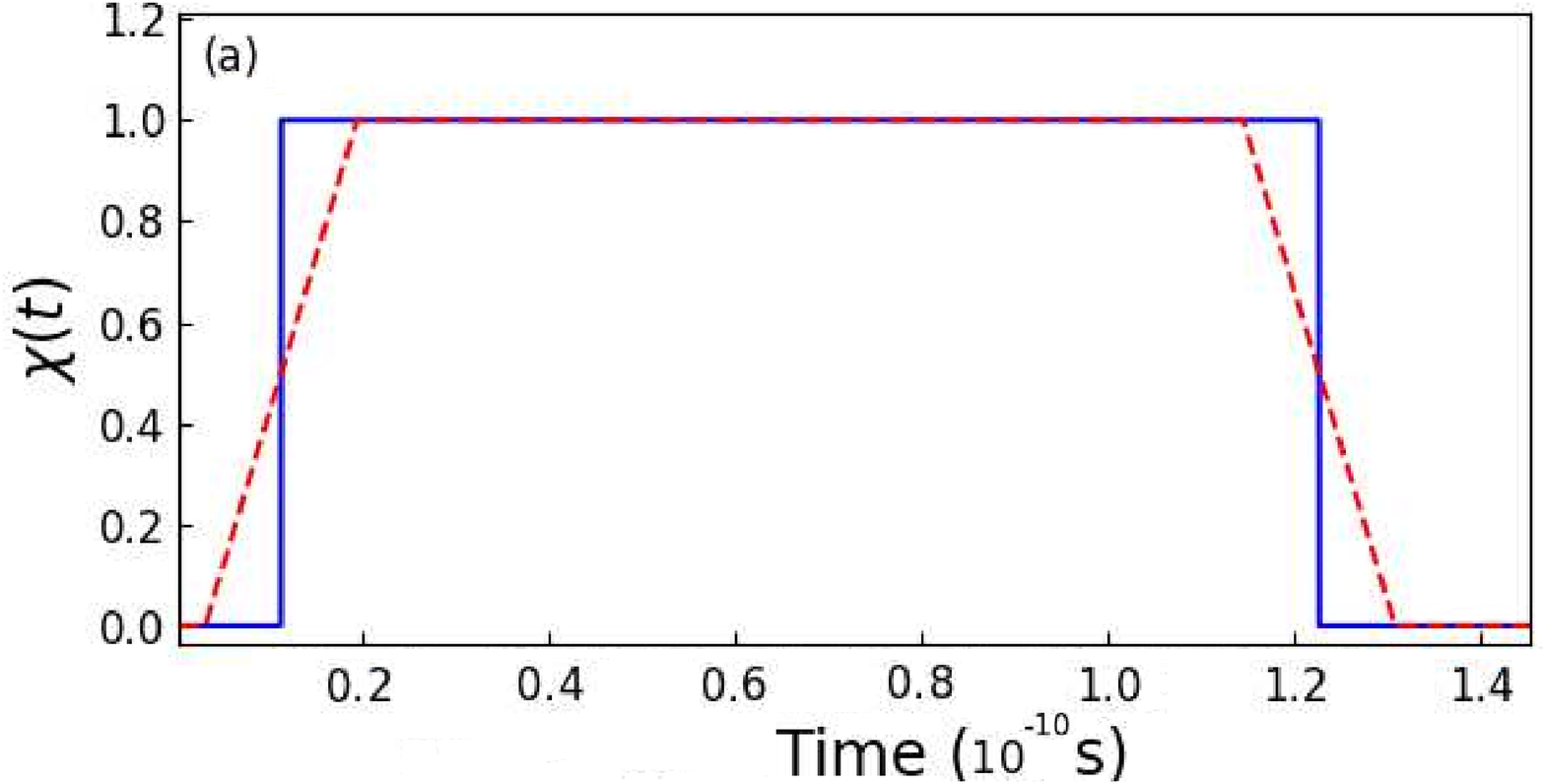}
\includegraphics[width=0.9\columnwidth]{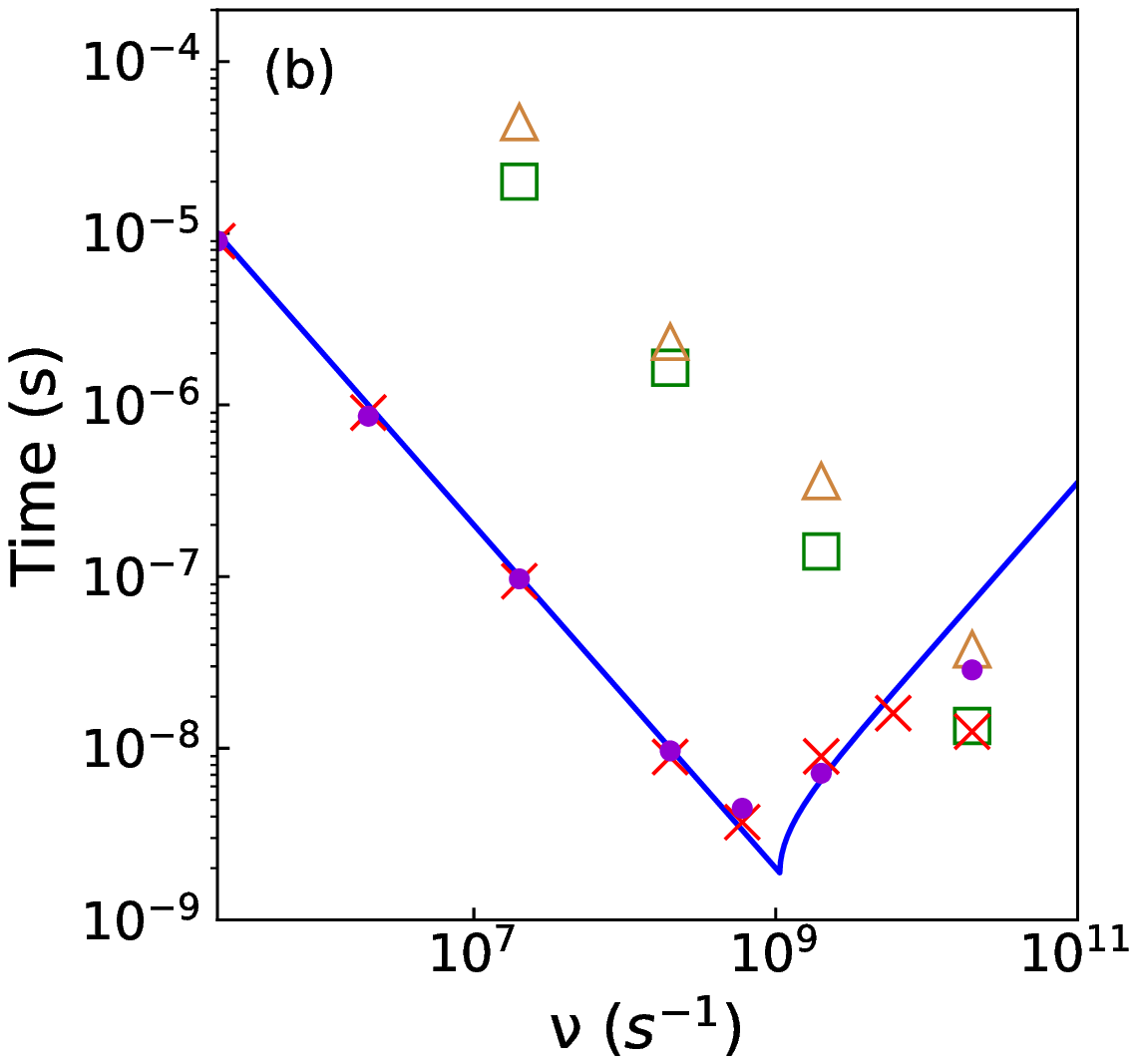}
\caption{(a) Modified telegraphic signal $\chi'(t)$ in which the transition between states 0 and 1 is linear over a time $\Delta t = 7$~ps (magenta dashed line), compared to the original telegraph signal $\chi(t)$ (blue solid line). (b) Characteristic lifetimes $T_1$ and $T_2$ versus tunneling rate $\nu$ calculated in the multi-level model $(N=10)$ for Trap 1. Green squares and brown triangles: $T_{1}'$ calculated using $\chi(t)$ or $\chi'(t)$, respectively. Red crosses and violet dots: $T_2$ calculated using $\chi(t)$ or $\chi'(t)$, respectively. The blue solid line depicts the analytical expression for $T_{2}^{*}$, as given by Eq.~(\ref{eq_anal_T2}), using $\omega_{th}$ and $|u_{\uparrow\downarrow}|$ of Table~\ref{table_param}.}
\label{fig_diabatic}
\end{figure}

We consider a modified telegraphic signal $\chi'(t)$ in which the fluctuator is assumed to vary progressively (linearly) between states 0 and 1 over a time $\Delta t = 7$~ps. Figure~\ref{fig_diabatic}(b) shows that the characteristic times calculated using $\chi'(t)$ behave as a function of $\nu$ in the same way as for the original telegraph signal $\chi(t)$. At low frequencies, $T_2$ remains given by $2/\nu$, the dephasing time remains limited by the average switching time of the fluctuator. On the other hand, $T_{1}'$ reaches higher values due to the fact that transitions to higher energy hole states are less likely. However, the overall behavior remains the same.

The question is therefore whether a value $\Delta t$ of 7~ps is realistic. This does not appear to be the case, as tunneling times are typically in the femtosecond range \cite{Landauer94,Fevrier18}, as can be estimated with the expression $\tau_{T}= d\sqrt{m/(2U_{b})}$, in which $d$ is the length of the tunneling barrier ($\approx 1$~nm), $U_b$ is its height ($\approx 2$~eV) and $m$ is the carrier effective mass ($\approx$ free electron mass). The characteristic times calculated for $\Delta t$ in the femtosecond range are those presented in Fig.~\ref{fig_T1_T2_multi_level}. Therefore, the instantaneous transitions model employed in this work seems justified.

\bibliography{paper_SI}

\end{document}